\documentclass[12pt,notitlepage]{iopart}
\usepackage{iopams,amsfonts,amssymb,latexsym,epsf,bm} 
\expandafter\let\csname equation*\endcsname\relax
\expandafter\let\csname endequation*\endcsname\relax
\usepackage{amsmath}
\usepackage[pdftex]{graphicx,color}
\newcommand{\bra}[1]{\langle #1|}
\newcommand{\ket}[1]{|#1 \rangle}

\newcommand{\corr}[1]{\langle #1 \rangle}

\newcommand{\av}{{\mathbf{a}}}
\newcommand{\bv}{{\mathbf{b}}}
\newcommand{\ev}{{\mathbf{e}}}
\newcommand{\rv}{{\mathbf{r}}}

\newcommand{\kv}{{\mathbf{k}}}
\newcommand{\pv}{{\mathbf{p}}}
\newcommand{\qv}{{\mathbf{q}}}
\newcommand{\mv}{{\mathbf{m}}}
\newcommand{\nv}{{\mathbf{n}}}

\newcommand{\uv}{{\mathbf{u}}}
\newcommand{\vv}{{\mathbf{v}}}

\newcommand{\Xv}{{\mathbf{X}}}

\newcommand{\zerov}{{\mathbf{0}}}

\newcommand{\Av}{{\mathbf{A}}}

\newcommand{\Kv}{{\mathbf{K}}}
\newcommand{\Gv}{{\mathbf{G}}}

\newcommand{\psih}{\hat{\psi}}

\newcommand{\bt}{\tilde{b}}
\newcommand{\Vt}{\tilde{V}}

\newcommand{\psit}{\tilde{\psi}}

\newcommand{\Ecal}{\mathcal{E}}
\newcommand{\Lcal}{\mathcal{L}}
\newcommand{\Ocal}{\mathcal{O}}
\newcommand{\Mcal}{\mathcal{M}}

\newcommand{\Zbb}{\mathbb{Z}}

\newcommand{\ua}{\uparrow}
\newcommand{\da}{\downarrow}

\newcommand{\GS}{\mathrm{GS}}

\newcommand{\Nvor}{N_\mathrm{v}}
\newcommand{\Nvoro}{N_\mathrm{v1}}
\newcommand{\Nvort}{N_\mathrm{v2}}

\newcommand{\thetas}{\theta_\mathrm{sing}}
\newcommand{\jvor}{j_\mathrm{vor}}
\newcommand{\sing}{\mathrm{sing}}
\newcommand{\reg}{\mathrm{reg}}

\newcommand{\elastic}{\mathrm{el}}

\newcommand{\Ucal}{{\cal U}}
\newcommand{\Vcal}{{\cal V}}
\newcommand{\erm}{{\mathrm{e}}}

\newcommand{\osc}{{\mathrm{osc}}}
\newcommand{\zero}{\mathrm{zero}}
\newcommand{\drm}{\mathrm{d}}
\newcommand{\irm}{\mathrm{i}}
\newcommand{\srm}{\mathrm{s}}
\newcommand{\Brm}{\mathrm{B}}
\newcommand{\gb}{\bar{g}}

\newcommand{\arsinh}{\mathrm{arsinh}}

\begin{document}
%%%%%%%%%%%%%%%%%%%%%%%%%%%%%%%%%%%%%%%%%%%%%%%%%
% Paper Information (revtex)
%%%%%%%%%%%%%%%%%%%%%%%%%%%%%%%%%%%%%%%%%%%%%%%%%
%\title{Vortex lattices in binary Bose-Einstein condensates: Collective modes, quantum fluctuations, and intercomponent entanglement}
%\author{Takumi Yoshino}
%\email{yoshino@cat.phys.s.u-tokyo.ac.jp}
%\affiliation{Department of Physics, University of Tokyo, 7-3-1 Hongo, Bunkyo-ku, Tokyo 113-0033, Japan}
%\author{Shunsuke Furukawa}
%\email{furukawa@rk.phys.keio.ac.jp}
%\affiliation{Department of Physics, %Faculty of Science and Technology, 
%Keio University, 3-14-1 Hiyoshi, Kohoku-ku, Yokohama 223-8522, Japan}
%\author{Masahito Ueda}
%\affiliation{Department of Physics, University of Tokyo, 7-3-1 Hongo, Bunkyo-ku, Tokyo 113-0033, Japan}
%\affiliation{RIKEN Center for Emergent Matter Science (CEMS), Wako, Saitama 351-0198, Japan}
%\affiliation{Institute for Physics of Intelligence, University of Tokyo, 7-3-1 Hongo, Bunkyo-ku, Tokyo 113-0033, Japan}
%\date{\today}

%%%%%%%%%%%%%%%%%%%%%%%%%%%%%%%%%%%%%%%%%%%%%%%%%
% Paper Information (IOP)
%%%%%%%%%%%%%%%%%%%%%%%%%%%%%%%%%%%%%%%%%%%%%%%%%

\title[]{Vortex lattices in binary Bose-Einstein condensates: Collective modes, quantum fluctuations, and intercomponent entanglement}
\author{Takumi Yoshino$^1$, Shunsuke Furukawa$^2$, Masahito Ueda$^{1,3,4}$}
\address{
$^1$Department of Physics, University of Tokyo, 7-3-1 Hongo, Bunkyo-ku, Tokyo 113-0033, Japan\\
$^2$Department of Physics, %Faculty of Science and Technology, 
Keio University, 3-14-1 Hiyoshi, Kohoku-ku, Yokohama 223-8522, Japan\\
$^3$RIKEN Center for Emergent Matter Science (CEMS), Wako, Saitama 351-0198, Japan\\
$^4$Institute for Physics of Intelligence, University of Tokyo, 7-3-1 Hongo, Bunkyo-ku, Tokyo 113-0033, Japan
}
\ead{yoshino@cat.phys.s.u-tokyo.ac.jp and furukawa@rk.phys.keio.ac.jp}

\vspace{1pc}

%%%%%%%%%%%%%%%%%%%%%%%%%%%%%%%%%%%%%%%%%%%%%%%%%
% Abstract
%%%%%%%%%%%%%%%%%%%%%%%%%%%%%%%%%%%%%%%%%%%%%%%%%

\begin{abstract}
We study binary Bose-Einstein condensates subject to synthetic magnetic fields in mutually parallel or antiparallel directions. 
Within the mean-field theory, the two types of fields have been shown to give the same vortex-lattice phase diagram. 
We develop an improved effective field theory to study properties of collective modes and ground-state intercomponent entanglement. 
Here, we point out the importance of introducing renormalized coupling constants for coarse-grained densities. 
We show that the low-energy excitation spectra for the two %types 
kinds
of fields are related to each other by suitable rescaling using the renormalized constants. 
By calculating the entanglement entropy, we find that 
for an intercomponent repulsion (attraction), the two components are more strongly entangled in the case of parallel (antiparallel) fields, 
in qualitative agreement with recent studies for a quantum (spin) Hall regime. 
We also find that the entanglement spectrum exhibits an anomalous square-root dispersion relation, 
which leads to a subleading logarithmic term in the entanglement entropy. 
All of these are confirmed by numerical calculations based on the Bogoliubov theory with the lowest-Landau-level approximation. 
Finally, we investigate the effects of quantum fluctuations on the phase diagrams 
by calculating the correction to the ground-state energy due to zero-point fluctuations in the Bogoliubov theory. 
We find that the boundaries between rhombic-, square-, and rectangular-lattice phases shift appreciably with a decrease in the filling factor. 
\end{abstract}
\maketitle

%%%%%%%%%%%%%%%%%%%%%%%%%%%%%%%%%%%%%%%%%%%%%%%%%
% Main text
%%%%%%%%%%%%%%%%%%%%%%%%%%%%%%%%%%%%%%%%%%%%%%%%%

%%%%%%%%%%%%%%%%%%%%%%%%%%%%%%%%%%%%%%%%%%%%%%%%%
\section{Introduction}
%%%%%%%%%%%%%%%%%%%%%%%%%%%%%%%%%%%%%%%%%%%%%%%%%

% [ Synthetic gauge fields ] --------------------------------------
Engineering synthetic gauge fields and observing their physical effects in ultracold atomic gases 
have been a subject of great interest in recent years \cite{Dalibard11, Goldman14, ZhangSL17, Aidelsburger18,Galitski19}. 
While atomic gases are charge neutral, effective gauge fields can be induced by rotating gases \cite{Cooper08, Fetter09} or optically dressing atoms \cite{Lin09}. 
Atomic Bose-Einstein condensates (BECs) in synthetic magnetic fields have close analogy with type-II superconductors in magnetic fields. 
Real or synthetic magnetic fields induce quantized vortices in these systems; when vortices proliferate in high fields, 
they %organize into a regular lattice 
form a regular lattice pattern 
owing to their %mutual 
repulsion, as originally predicted by Abrikosov \cite{Abrikosov56}. 
The resulting triangular vortex lattice structure has been observed experimentally in rapidly rotating BECs \cite{AboShaeer01,Engels02,Schweikhard04_1comp}. 
A vortex lattice %supports an elliptically polarized oscillatory mode 
exhibits elliptically polarized oscillations 
known as the Tkachenko mode \cite{Tkachenko66a,Tkachenko66b,Tkachenko69,Sonin87}, 
which has also been observed in a trapped BEC \cite{Schweikhard04_1comp,Coddington03}. 
In a uniform system, the Tkachenko mode has a quadratic dispersion relation 
at low frequencies \cite{Sinova02,Baym03,Baym04,Matveenko11,Kwasigroch12}, 
and is understood as a Nambu-Goldstone mode associated with 
spontaneously broken U(1) symmetry and magnetic translation symmetries \cite{WatanabeMurayama13}. 
For sufficiently high synthetic magnetic fields, 
%it is reasonable to assume that atoms reside 
atoms are expected to be 
in the lowest Landau level. 
A key parameter in this regime is the filling factor $\nu:=N/\Nvor$, 
where $N$ is the number of atoms and $\Nvor$ is the number of flux quanta piercing the system. 
While the Gross-Pitaevskii (GP) mean-field theory is applicable for $\nu\gg 1$ \cite{Ho01}, 
quantum fluctuations become significant as $\nu$ is lowered \cite{Sinova02,Baym04,Matveenko11,Kwasigroch12,Sonin05}. 
Theory predicts that when $\nu$ is below a critical value $\nu_\mathrm{c}$, the vortex lattice melts  
and incompressible quantum Hall states appear 
at various integer and fractional values of $\nu$ \cite{Cooper08, Wilkin98, Cooper01,Cooper20}. 
Estimates of $\nu_\mathrm{c}$ vary between $\nu_\mathrm{c}\simeq 2-6$ from exact diagonalization \cite{Cooper01,Cooper07,Liu11}
and $\nu_\mathrm{c}\simeq 5-14$ from a Lindemann criterion \cite{Sinova02,Baym04,Kwasigroch12,Cooper01}.

\begin{figure*}
\centering\includegraphics[width=16cm, angle=0]{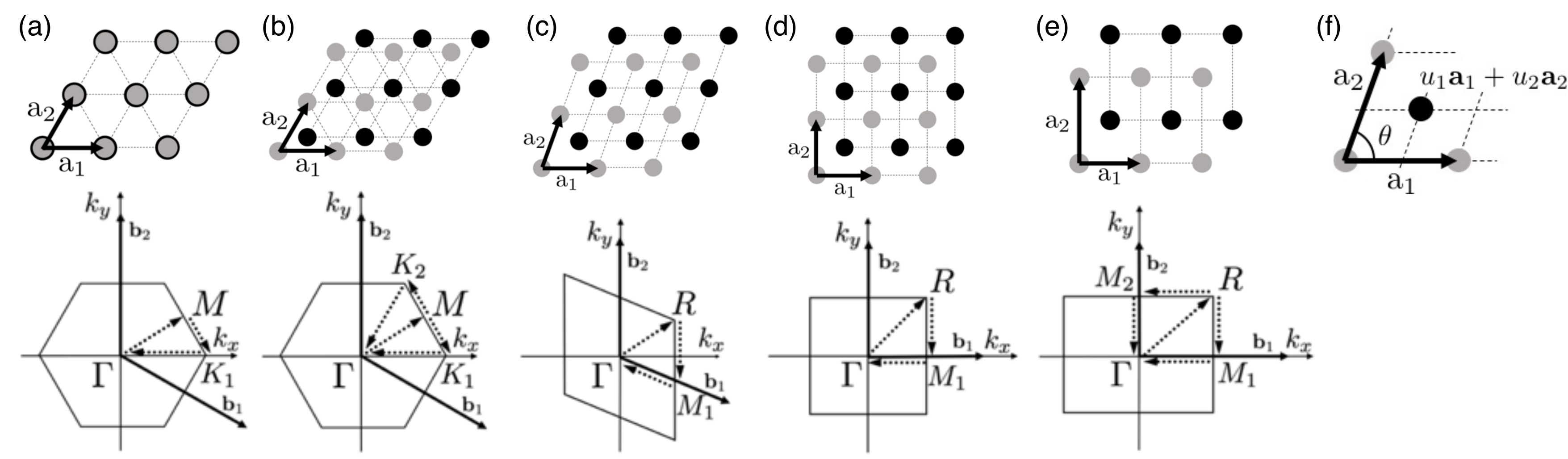}
\caption{\label{fig:Phases_GP}
Upper panels: vortex-lattice structures in the ground state of the binary BECs 
under synthetic magnetic fields described by the Lagrangian in Eq.\ \eqref{eq:2comp_Lag} 
\cite{Mueller02,Kasamatsu03,Kasamatsu05,Mingarelli18} (see also Refs.\ \cite{Woo07,Barnett08,Mason11,Aftalion12,Kuopanportti12,Kumar17,Mingarelli19}). 
Within the GP mean-field theory, the systems in parallel and antiparallel fields exhibit the same phase diagrams \cite{Furukawa14}. 
Five different structures appear as
%the ratio of the intercomponent coupling $g_{\ua\da}$ to the intracomponent one $g$ 
the interaction ratio $g_{\ua\da}/g$ 
is varied (see Fig.\ \ref{fig:GS energy}): 
(a) overlapping triangular lattices, (b) interlaced triangular lattices, (c) rhombic lattices, 
(d) square lattices, and (e) rectangular lattices. 
%We assume that the same types of structures appear when we analyze the effects of quantum fluctuations on the phase diagrams in Sec.\ \ref{sec:phase_diagrams}. 
Black (grey) circles %indicate 
show 
the vortex positions in the spin-$\ua$ ($\da$) component. 
As shown in (f), each lattice structure is %characterized 
specified 
by the primitive vectors $\av_1=(a,0)$ and $\av_2=b(\cos\theta,\sin\theta)$ 
as well as the vortex displacement $u_1\av_1+u_2\av_2$ in one component relative to the other. 
The area of the unit cell is given by $(\av_1\times\av_2)_z=ab\sin\theta=2\pi\ell^2$. 
The angle $\theta$ and the aspect ratio $b/a$ vary continuously in the rhombic- and rectangular-lattice phases, respectively, 
as shown later in Fig.\ \ref{fig:parameters}. 
Lower panels: the %first 
Brillouin zone for each lattice structure %placed 
shown 
above. 
%The reciprocal primitive vectors are given by $\bv_1=(b\sin\theta,-b\cos\theta)/\ell^2$ and $\bv_2 = (0,a)/\ell^2$. 
The vectors $\bv_1$ and $\bv_2$ show the reciprocal primitive vectors, while
uppercase letters %indicate 
show 
high-symmetry points. 
The excitation spectra and the single-particle ES presented in Figs.~\ref{fig:spec_resc} and \ref{fig:ES_GP} 
are calculated along %the paths indicated by 
dotted arrows.
}
\end{figure*}

% [ Binary BECs ] --------------------------------------
For binary (or pseudospin-$\frac12$) BECs, which are populated in two hyperfine spin states of the same atomic species, 
a richer variety of synthetic gauge fields have been realized, such as a uniform magnetic field by rotation \cite{Schweikhard04_2comp}, 
and spin-orbit couplings \cite{Lin11,Zhai12_review, WuZ16} and pseudospin-dependent antiparallel magnetic fields \cite{Beeler13} by optical dressing techniques. 
For binary BECs under rotation, GP mean-field calculations have %shown 
revealed 
that five %types of vortex lattices 
vortex-lattice phases 
appear as the ratio of the intercomponent coupling $g_{\ua\da}$ to the intracomponent one $g>0$ is varied 
(see Fig.\ \ref{fig:Phases_GP}) \cite{Mueller02,Kasamatsu03,Kasamatsu05,Mingarelli18}. 
%Among them, interlaced 
Square vortex lattices [Fig.\ \ref{fig:Phases_GP}(d)]
%, which are unique to binary systems, 
have indeed been observed experimentally \cite{Schweikhard04_2comp}. 
Meanwhile, a spin Hall effect due to pseudospin-dependent Lorentz forces has been observed in binary BECs in antiparallel magnetic fields \cite{Beeler13}. 
%If the antiparallel fields are made even higher, such systems are expected to show 
For high antiparallel fields, theory predicts 
a rich phase diagram 
that consists of vortex lattices and (fractional) quantum spin Hall states \cite{Liu07,Fialko14,Furukawa14}. 
%Notably, it has been shown {\it within} the GP mean-field theory that 
Remarkably, {\it within} the GP mean-field theory, 
binary BECs in antiparallel magnetic fields 
exhibit the {\it same} %vortex-lattice 
vortex 
phase diagram as those in parallel magnetic fields (i.e., under rotation) \cite{Furukawa14}. 
This is because the GP energy functionals as well as the time-independent GP equations for the two systems are related to each other 
through the complex conjugation of the spin-$\da$ condensate wave function. 
It is interesting to further investigate similarities and differences between the two systems. 
As the systems in parallel and antiparallel magnetic fields are closely related 
with bilayer quantum Hall systems \cite{Girvin97} and quantum spin Hall systems \cite{Bernevig06}, respectively, 
a comparative study of these systems can make a link between the two research fields. 
Such studies have been conducted on 
collective modes of vortex lattices \cite{Oktel06,Yoshino19} 
and phase diagrams in a quantum (spin) Hall regime \cite{Furukawa14,Furukawa13,WuYH13,Regnault13,WuYH15,Geraedts17,Furukawa17}. 

%############################
%############################

% [ Collective modes ] --------------------------------------
Collective excitation spectra can be different between the parallel- and antiparallel-field cases 
as there is no correspondence between the two cases for the time-dependent GP equations. 
Ke\c{c}eli and Oktel \cite{Oktel06} have calculated excitation spectra in binary BECs %in parallel fields by means of 
under rotation by using 
a hydrodynamic theory. 
In a previous work \cite{Yoshino19}, we have conducted a comparative study of excitation modes 
in the parallel- and antiparallel-field cases by means of the Bogoliubov theory and an effective field theory. 
All the calculations in these works are based on the lowest-Landau-level (LLL) approximation. 
For both types of fields and for all the lattice structures in Fig.\ \ref{fig:Phases_GP}, 
it is found that there appear two distinct modes with linear and quadratic dispersion relations at low energies, 
which exhibit anisotropy reflecting the symmetry of each lattice structure. 
Furthermore, for overlapping triangular lattices [Fig.\ \ref{fig:Phases_GP}(a)], 
the low-energy spectra for the two types of fields are found to be related to each other by simple rescaling \cite{Yoshino19}. 
This indicates a nontrivial correspondence between the two types of fields in excitation properties. 
However, such rescaling relations 
do not hold for other lattices, which appears inconsistent with the effective field theory prediction. 
A more refined description of low-energy modes has remained an open issue. 

% [ Quantum Hall regime ] --------------------------------------
Numerical studies on the quantum (spin) Hall regime with $\nu=O(1)$
have revealed that binary Bose gases in parallel and antiparallel synthetic magnetic fields
exhibit markedly different phase diagrams \cite{Furukawa14,Furukawa13,WuYH13,Regnault13,WuYH15,Geraedts17,Furukawa17}. 
For parallel fields, the product states of a pair of nearly uncorrelated quantum Hall states
are robust against an intercomponent attraction $g_{\ua\da}<0$ 
and persist even when $|g_{\ua\da}|$ is close to $g$ \cite{Furukawa17}. 
Meanwhile, a variety of spin-singlet quantum Hall states with high intercomponent entanglement emerge 
for $g_{\ua\da}\approx g$ \cite{Furukawa13,WuYH13,Regnault13,WuYH15,Geraedts17}. 
For antiparallel fields, (fractional) quantum spin Hall states approximated by products of quantum Hall states with opposite chiralities 
are robust against an intercomponent repulsion $g_{\ua\da}>0$ \cite{Furukawa14}. 
The phase diagrams for the two types of fields thus exhibit opposite behavior in view of intercomponent entanglement. 
An interpretation of these results has been given in light of pseudopotential representations of interactions \cite{Furukawa17}. 
As there is no intercomponent entanglement in the GP mean-field theory valid for $\nu\gg 1$, 
it is interesting to investigate how the intercomponent entanglement arises in the two systems as $\nu$ is lowered from the mean-field regime. 

% [ This paper ]--------------------------------------
In this paper, we present a detailed comparative study of vortex lattices of binary BECs in parallel and antiparallel fields 
concerning ground-state and excitation properties. 
We first formulate an improved effective field theory for such vortex lattices, and derive some properties of collective modes and ground-state intercomponent entanglement. 
Here, a major improvement is the introduction of renormalized coupling constants for coarse-grained densities.\footnote{
In the formulation of Ref.\ \cite{Yoshino19}, the necessity of renormalization was not transparent. 
This was because we integrated out the density variables, before coarse graining, 
to obtain an effective Lagrangian for phase variables. 
In this paper, we keep both the phase and density variables in the Lagrangian, and perform coarse graining of both the variables at the same time.  
The renormalization of density-density interactions can naturally be understood in this formulation. 
This formulation also allows us to move easily to the operator formalism, in which ground-state and excitation properties can be studied in an algebraic manner. 
}
We show that the low-energy excitation spectra for the two types of fields are related to each other by suitable rescaling using the renormalized constants. 
Namely, the rescaling relations proposed in our previous work \cite{Yoshino19} must be modified using the renormalized constants. 
By calculating the entanglement entropy (EE), we find that for an intercomponent repulsion (attraction), 
the two components are more strongly entangled in the case of parallel (antiparallel) fields, 
in qualitative agreement with recent exact diagonalization results for a quantum (spin) Hall regime \cite{Furukawa14,Furukawa17}. 
As a by-product, we also find that the entanglement spectrum (ES) exhibits an anomalous square-root dispersion relation, 
and that the EE exhibits a volume-law scaling with a subleading logarithmic correction. 
%which leads to a subleading logarithmic term in the EE. 
This anomalous feature of the ES is associated with the emergence of a long-range interaction in terms of the density in the entanglement Hamiltonian. 
All these predictions are confirmed by numerical calculations based on the Bogoliubov theory with the LLL approximation. 
Finally, we investigate the effects of quantum fluctuations on the phase diagrams 
by calculating the Lee-Huang-Yang correction, which is a correction to the ground-state energy due to zero-point fluctuations 
in the Bogoliubov theory \cite{LeeHuangYang57}. 
We find that the boundaries between rhombic-, square-, and rectangular-lattice phases shift appreciably with a decrease in $\nu$. 
Here, the shift occurs more significantly for parallel fields. 

% [ Previous work on intercomponent entanglement ]--------------------------------------
Let us comment on the relation to another recent work of our own \cite{Yoshino21}
concerning intercomponent ES and EE in binary BECs in $d$ spatial dimensions 
in the absence of synthetic gauge fields. 
Here we employ effective field theory to show that 
the ES exhibits a gapless square-root dispersion relation in the presence of an intercomponent tunneling (a Rabi coupling) 
and a gapped dispersion relation in its absence (see also Refs.\ \cite{Chen13,Lundgren13} for related results in two coupled Tomonaga-Luttinger liquids). 
%In the former case, we have also found that the EE exhibits a volume-law scaling 
%with a subleading logarithmic term $\frac{d-1}{2} \ln L$, where $L$ is the linear system size [see Eq.\ (49) in Ref.\ \cite{Yoshino19}]. 
%In the coefficient $(d-1)/2$, $d/2$ originates from the symmetry restoration for a finite volume and $-1/2$ from the Nambu-Goldstone modes. 
In the present work, in contrast, the ES exhibits a square-root dispersion relation 
in the absence of an intercomponent tunneling in both the parallel- and antiparallel-field cases. 
This qualitative distinction is related to the fact that binary BECs in parallel and antiparallel fields 
have a higher density of low-energy excitations and thus experience larger quantum fluctuations than those without synthetic gauge fields. 
References \cite{Yoshino21,Lundgren13} and the present paper demonstrate that 
a variety of long-range interactions can be emulated in a subsystem of multicomponent BECs that have only short-range interactions. 
We also note that the field-theoretical methods for investigating intercomponent entanglement in the present paper 
are closely analogous to those in Refs.\ \cite{Yoshino21,Chen13,Lundgren13}. 

% [ Organization of the paper ]--------------------------------------
The rest of this paper is organized as follows. 
In Sec.\ \ref{sec:EFT_vorlat}, we introduce the systems that we study in this paper, 
and formulate the low-energy effective field theory to derive some properties of excitation spectra, intercomponent entanglement, and correlation functions. 
In Sec.\ \ref{sec:BogoliubovTheory}, we briefly review the Bogoliubov theory with the LLL approximation, 
which has been adapted to the present problem in Ref.\ \cite{Yoshino19}, 
and then give the expressions of the intercomponent ES and EE. 
In Sec.\ \ref{sec:NumericalResults}, we present numerical results based on the Bogoliubov theory. 
We confirm the field-theoretical predictions on the excitation spectra, the intercomponent entanglement, and the fraction of quantum depletion. 
Furthermore, we investigate the effects of quantum fluctuations on the ground-state phase diagrams. 
%by calculating the correction to the ground-state energy in the Bogoliubov theory. 
In Sec.\ \ref{sec:Summary}, we present a summary of the present study and an outlook for future studies. 
In Appendices, we describe some technical details of Secs.\ \ref{sec:EFT_vorlat}.

%%%%%%%%%%%%%%%%%%%%%%%%%%%%%%%%%%%%%%%%%%%%%%%%
\section{Effective field theory}\label{sec:EFT_vorlat}
%%%%%%%%%%%%%%%%%%%%%%%%%%%%%%%%%%%%%%%%%%%%%%%%

Effective field theory for a vortex lattice in a scalar BEC has been developed in Refs.\ \cite{Sinova02,WatanabeMurayama13,Moroz18,Moroz19}. 
In our previous work \cite{Yoshino19}, we applied the formulation by Watanabe and Murayama \cite{WatanabeMurayama13} 
to binary BECs in parallel and antiparallel fields. 
For parallel fields, this approach was essentially equivalent to the hydrodynamic theory by Ke\c{c}eli and Oktel \cite{Oktel06}. 
Here, we formulate an improved effective field theory for binary BECs in parallel and antiparallel fields 
by introducing renormalized coupling constants $\gb_{\alpha\beta}$ for coarse-grained densities. 
In doing so, we keep both the phase and density variables in the Lagrangian, rather than integrating out the density variables as in Ref.\ \cite{Yoshino19}.
We then move to the operator formalism, and study excitation spectra, intercomponent entanglement, and correlation functions in an algebraic manner 
(see Refs.\ \cite{Yoshino21,Chen13,Lundgren13} for analogous calculations in different systems).  

%While the formulation of Ref.\ \cite{Yoshino19} was based on an effective Lagrangian for phase variables, 
%we here write down an effective Hamiltonian in terms of the density and phase operators, which are conjugate to each other. 
%This Hamiltonian formulation is useful for the analysis of correlation and entanglement properties in the ground state 
%as the ground state is expressed simply as the vacuum of elementary excitations 

%************************************************
\subsection{Systems}
%************************************************

% [ Systems ]--------------------------------------
We consider a system of 2D binary (pseudospin-$\frac12$) BECs having two hyperfine spin states (labeled by $\alpha=\ua, \da$)  
and subject to synthetic magnetic fields $B_\ua$ and $B_\da$ in mutually parallel or antiparallel directions. 
The Lagrangian density of the system is given by
\begin{equation}\label{eq:2comp_Lag}
 \mathcal{L}
 =\sum_{\alpha=\ua,\da}
 \left[\frac{i\hbar}{2}(\psi_\alpha^*\dot{\psi}_\alpha - \dot{\psi}_\alpha^*\psi_\alpha)
  -\frac{1}{2M}|(-i \hbar\nabla- q\mathbf{A}_\alpha)\psi_\alpha |^2\right]
  -\sum_{\alpha,\beta=\ua,\da}\frac{g_{\alpha\beta}}{2}|\psi_\alpha|^2|\psi_\beta|^2 ,
\end{equation}
where $\psi_\alpha(\rv,t)$ is the bosonic field for the spin-$\alpha$ component (with $\rv=(x,y)$ being the 2D coordinate) 
and $M$ and $q$ are the mass and the fictitious charge of an atom. 
The gauge field $\mathbf{A}_\alpha(\rv)$ for %the spin-$\alpha$ component 
spin-$\alpha$ bosons 
is given by 
\begin{equation}
 \mathbf{A}_\alpha =\frac{B_\alpha}{2}\ev_z\times\rv =\epsilon_\alpha\frac{B}{2}(-y,x) ,
\end{equation}
where we assume $qB>0$ and $\epsilon_\ua=\epsilon_\da=1$ $(\epsilon_\ua=-\epsilon_\da=1)$ for parallel (antiparallel) fields. 
%For a 2D system of area $A$, 
The number of magnetic flux quanta piercing each component 
%(i.e., the number of vortices) 
is given by $\Nvor =qBA/(2\pi\hbar)=A/(2\pi\ell^2)$, 
where $A$ is the area of the system and $\ell:=\sqrt{\hbar/qB}$ is the magnetic length. 
In the Lagrangian \eqref{eq:2comp_Lag}, the numbers of spin-$\ua$ and $\da$ atoms,  $N_\ua$ and $N_\da$, are separately conserved. 
We introduce the total filling factor $\nu:=N/\Nvor$, where $N:=N_\ua+N_\da$ is the total number of atoms. 

% [ Interactions ]--------------------------------------
We assume a contact interaction between atoms. 
For simplicity, we set $g_{\ua\ua}=g_{\da\da}\equiv g >|g_{\ua\da}|$ and $N_\ua=N_\da$ in the following. 
With these conditions, the system in parallel fields is invariant under the interchange of the two components, 
while the system in antiparallel fields is invariant under time reversal. 
To apply the LLL approximation, we further assume that 
%the synthetic magnetic fields $B_\alpha$ are sufficiently high or the interactions are sufficiently weak so that 
the scale of the interaction energy per atom, $|g_{\alpha\beta}| n$, is much smaller than the Landau-level spacing $\hbar\omega_\mathrm{c}$. 
Here, $n:=N_\ua/A=N_\da/A$ is the average density of $\ua$ or $\da$ atoms, and $\omega_\mathrm{c}:=qB/M$ is the cyclotron frequency. 

%For two dimensional system where a gas is tightly confined in a harmonic potential with frequency $\omega_z$ in the $z$ direction, 
%the effective interaction parameters are given by 
%$g_{\alpha\alpha}=a_\alpha \sqrt{8\pi\hbar^3\omega_z/M}$ and $g_{\ua\da}=g_{\da\ua}=a_{\ua\da}\sqrt{8\pi\hbar^3\omega_z/M}$, 
%where $a_\alpha$ and $a_{\ua\da}$ are $s$-wave scattering lengths between like and unlike bosons, respectively, in the three dimensional space. 

%************************************************
\subsection{Effective field theory}\label{sec:EFT_vorlat_derive}
%************************************************

%In the following, we present the effective Hamiltonian formulation of the problem, and use it to calculate the entanglement spectra. 

% [ Density and phase variables ]--------------------------------------
To obtain a low-energy effective description, it is useful to rewrite the field as
$\psi_{\alpha} = \erm^{-\irm \theta_{\alpha}} \sqrt{n_{\alpha}}$, 
where $n_{\alpha} (\rv,t)$ and $\theta_{\alpha}(\rv,t)$ are the density and phase variables, respectively. 
The Lagrangian density \eqref{eq:2comp_Lag} is rewritten in terms of these variables as 
\begin{equation}\label{eq:Lag_theta_n_pre}
 \mathcal{L}
 =\sum_{\alpha} \left[\hbar n_\alpha\dot{\theta}_{\alpha}
 -\frac{n_\alpha}{2M}(\hbar\nabla\theta_{\alpha}+q\mathbf{A}_{\alpha})^2
 -\frac{\hbar^2 (\nabla n_\alpha)^2}{8n_\alpha M} \right]
 -\sum_{\alpha,\beta} \frac{g_{\alpha\beta}}{2} n_\alpha n _\beta.
\end{equation}

% [ Decomposition into regular and singular parts ]--------------------------------------
In the presence of vortices, 
the phase variables $\{ \theta_\alpha(\rv,t) \}$ involve singularities. 
%It is therefore useful 
This motivates us 
to decompose $\theta_{\alpha}$ into regular and singular contributions as 
$\theta_{\alpha} =\theta_{\mathrm{reg},\alpha} + \theta_{\mathrm{sing},\alpha}$.
We also introduce the displacement $\mathbf{u}_{\alpha}(\rv,t)$ of a vortex from the equilibrium position. 
% as $\mathbf{u}_{\alpha}(\rv,t)=\rv - \Xv_{\alpha}(\mathbf{r},t)$,
%where $\rv $ is the equilibrium position of the vortex and $\Xv_{\alpha}$ is the position at time $t$. 
The derivatives of the singular part $\theta_{\mathrm{sing},\alpha}$ of the phase 
can be related to the displacement field $\uv_\alpha$ as \cite{WatanabeMurayama13,Moroz18}
\begin{equation}
 \hbar\dot{\theta}_{\sing,\alpha}
 =-\frac{qB_\alpha}{2} (\uv_\alpha \times\dot{\uv}_\alpha)_z, ~~
 \hbar\nabla\theta_{\sing,\alpha}+q\Av_\alpha 
 = qB_\alpha\ev_z\times\uv_\alpha-\frac{qB_\alpha}{2}\epsilon_{ij}u^i_\alpha \nabla u^j_\alpha,
\end{equation}
where $\epsilon_{ij}$ is an antisymmetric tensor with $\epsilon_{xy}=-\epsilon_{yx}=+1$.
%One should also note that 
The displacement $\uv_\alpha(\rv,t)$ %leads to 
also results in 
a change 
in the elastic energy $\int \mathrm{d}^2\rv~\Ecal_\elastic(\uv_\alpha, \partial_i\uv_\alpha)$, 
whose explicit form will be given in Sec.\ \ref{sec:elastic}.
The Lagrangian density can then be expressed in terms of $\{\theta_{\mathrm{reg},\alpha}, \uv_\alpha, n_\alpha \}$ as
\begin{equation}\label{eq:Eff_Lag}
\begin{split}
 \mathcal{L}
 =&\sum_{\alpha}\Biggl[\hbar n_\alpha \dot{\theta}_{\reg,\alpha}
  -\frac{qB_{\alpha}n_\alpha}{2}(\uv_{\alpha}\times\dot{\uv}_{\alpha})_z
  -\frac{n_\alpha}{2M}\left(\hbar\nabla\theta_{\mathrm{reg},\alpha}
   +qB_{\alpha}\mathbf{e}_z\times\uv_\alpha
   -\frac{qB_{\alpha}}{2}\epsilon_{ij}u_{\alpha}^i\nabla u_{\alpha}^j\right)^2  \\
 &~~~~~
   -\frac{\hbar^2(\nabla n_\alpha)^2}{8n_\alpha M}\Biggr] 
 -\sum_{\alpha,\beta} \frac{\bar{g}_{\alpha\beta}}{2} n_\alpha n _\beta -\Ecal_\elastic.
\end{split}
\end{equation}
Henceforth, we ignore the term $-\frac{qB_{\alpha}}{2}\epsilon_{ij}u_{\alpha}^i\nabla u_{\alpha}^j$ in the round brackets 
as it only gives more than quadratic contributions to $\Lcal$ in terms of $\nabla \theta_{\mathrm{reg}, \alpha}$ and $\uv_\alpha$. 
We also omit the subscript ``reg'' in $\theta_{\reg,\alpha}$. 

% [ Renormalization of interactions ]--------------------------------------
In the effective Lagrangian density \eqref{eq:Eff_Lag}, we have introduced renormalized coupling constants $\bar{g}_{\alpha\beta}$ 
with $\bar{g}_{\ua\ua}=\bar{g}_{\da\da}\equiv \bar{g}$ and $\bar{g}_{\ua\da}=\bar{g}_{\da\ua}$. 
The necessity of this renormalization has been overlooked in previous studies \cite{Oktel06,Yoshino19} and can be explained as follows. 
The introduction of the vortex displacement fields $\{u_\alpha(\rv,t)\}$ necessarily involves the coarse graining of the theory. 
Namely, we smooth out details within the scale of the lattice constants and instead focus on the physics at larger scales. 
Therefore, the density variable $n_\alpha(\rv,t)$ should now be understood as the density averaged over the unit cell containing $\rv$. 
The renormalized coupling constants $\bar{g}_{\alpha\beta}$ can then be introduced through the relation
\begin{equation}\label{eq:g_renorm}
  \int_\mathrm{u.c.} \drm^2\rv'~g_{\alpha\beta} |\psi_\alpha(\rv',t)|^2 |\psi_\beta(\rv',t)|^2 = 2\pi\ell^2 \bar{g}_{\alpha\beta} n_\alpha(\rv,t) n_\beta(\rv,t),
\end{equation}
where $|\psi_\alpha|^2$ and $n_\alpha$ are the original and coarse-grained densities of the spin-$\alpha$ atoms, respectively, 
and the integration on the left-hand side is taken over the unit cell (with the area $2\pi\ell^2$) that contains $\rv$.
As explained later in this section and in Sec.\ \ref{sec:numerics_renorm}, the renormalized constants $\bar{g}_{\alpha\beta}$ for describing the low-energy physics 
can be determined by calculating the contribution of each interaction term to the mean-field ground-state energy. 
As seen in Eq.\ \eqref{eq:g_renorm}, a positive (negative) correlation between the density fluctuations $|\psi_\alpha|^2-n_\alpha$ and $|\psi_\beta|^2-n_\beta$ 
leads to an enhanced (reduced) renormalized coupling $|\bar{g}_{\alpha\beta}|>|g_{\alpha\beta}|$ ($|\bar{g}_{\alpha\beta}|<|g_{\alpha\beta}|$). 
In particular, the intracomponent coupling $g$ is always enhanced by the renormalization 
while the intercomponent repulsion $g_{\ua\da}>0$ (attraction $g_{\ua\da}<0$) is reduced (enhanced) by the renomalization 
owing to the displacement (overlap) of vortices between the components; this can be confirmed in Fig.\ \ref{fig:renorm_int} below. 
For a rotating scalar BEC with a repulsive coupling $g>0$, a similar renormalization with $\bar{g}/g=1.1596$ has been discussed in Refs.\ \cite{Fetter09,Sinova02,Aftalion05}. 
At high filling factors $\nu\gg 1$ and for not too large a number of vortices $\Nvor$, we can assume that the condensates are only weakly depleted; 
we therefore have $|n_\alpha(\rv,t)-n|\ll n$ for the coarse-grained density $n_\alpha$. 

%(as shown a posteriori in Sec.\ \ref{sec:EFT_vorlat_corr}); 

% [ Relation between the vortex displacement and the phase ]--------------------------------------
Because the displacement fields $\{\uv_\alpha\}$ involve the mass term $-\uv_\alpha^2$ in Eq.\ \eqref{eq:Eff_Lag}, 
%one can expect that they can safely be integrated out 
one may safely integrate them out 
in the discussion of low-energy dynamics. 
%To do so, 
Instead of performing the integration directly, 
it is useful to derive the Euler-Lagrange equations for $\{\uv_\alpha\}$:  
\begin{equation}\label{eq:EOM_u}
 \uv_\alpha - \epsilon_\alpha \ell^2 \ev_z\times\nabla\theta_\alpha
 - \frac{\epsilon_\alpha}{\omega_\mathrm{c}} \ev_z\times\dot{\uv}_\alpha
 +\frac{\ell^2}{n \hbar\omega_\mathrm{c}} 
 \left[\frac{\partial\Ecal_\elastic}{\partial \uv_\alpha} 
 -\partial_j\left(\frac{\partial\Ecal_\elastic}{\partial\left(\partial_j\uv_\alpha\right)}\right)\right] =0,
\end{equation}
where we have made the approximation $n_\alpha\approx n$. 
We can ignore 
the third and fourth terms on the left-hand side %can be ignored 
in the LLL approximation 
with $\hbar\omega, \Ecal_\elastic/n \ll \hbar\omega_\mathrm{c}$, 
where $\omega$ is the frequency of our %interest. 
concern. 
Introducing $\uv_\pm:=\uv_\ua\pm\uv_\da$ and $\theta_\pm:=\theta_\ua\pm\theta_\da$, 
we can rewrite Eq.\ \eqref{eq:EOM_u} %can be rewritten 
as
\begin{equation}\label{eq:EOM para/anti}
 \uv_{\pm} 
 =\left\{ \begin{aligned}
   &\ell^2 \mathbf{e}_z \times \nabla \theta_{\pm}  \qquad \text{(parallel fields)} ; \\
   &\ell^2 \mathbf{e}_z \times \nabla \theta_{\mp}  \qquad \text{(antiparallel fields)} .
 \end{aligned} \right. \qquad \text{}
\end{equation}
These relations indicate that the (anti)symmetric movement of vortices is coupled to the (anti)symmetric component of the phase variables in parallel fields 
while they are coupled in a crossed manner in antiparallel fields. 
By substituting Eq.\ \eqref{eq:EOM para/anti} into Eq.\ \eqref{eq:Eff_Lag}, we obtain the effective Lagrangian 
in terms of $\theta_\pm$ and their conjugate momenta $\hbar n_\pm=\hbar(n_\ua\pm n_\da)/2$. 
The Hamiltonian density is then obtained as
\begin{equation}\label{eq:Hcal_n_theta}
 \mathcal{H}=\sum_{\nu=\pm}\left[\gb_\nu n_\nu^2 + \frac{\hbar^2(\nabla n_\nu)^2}{4nM}\right]+\Ecal_\elastic,~~~\gb_\pm:=\bar{g} \pm \bar{g}_{\ua\da}.
\end{equation}
where $\Ecal_\elastic$ is expressed in terms of the phase variables $\theta_\pm$ by using Eq.\ \eqref{eq:EOM para/anti}. 
The theory can be quantized by requiring the canonical commutation relations 
\begin{equation}
 [ \theta_\nu(\rv) , n_{\nu'}(\rv') ] = \irm\delta_{\nu\nu'} \delta(\rv-\rv') ~~~(\nu,\nu'=\pm). 
\end{equation}

% [  ]--------------------------------------
In the present coarse-grained description, the mean-field ground state corresponds to the uniform state with 
$n_+(\rv)=n$, $n_-(\rv)=0$, and $\nabla \theta_\pm(\rv)=0$. 
Therefore, using Eq.\ \eqref{eq:Hcal_n_theta}, 
we obtain the mean-field ground-state energy density as 
\begin{equation}\label{eq:EGS_fieldtheory}
 \frac{E_\mathrm{GS}^\mathrm{MF}}{A}=\gb_+n^2=(\bar{g}+\bar{g}_{\ua\da})n^2 = \frac12 \sum_{\alpha,\beta} \gb_{\alpha\beta} n^2.
\end{equation}
In contrast, the same energy density obtained by the microscopic calculation [the first term of Eq.\ \eqref{eq:Egs/atom} shown later] 
has the form $E_\mathrm{GS}^\mathrm{MF}/A=(\beta g+\beta_{\ua\da} g_{\ua\da})n^2$, 
where $\beta$ and $\beta_{\ua\da}$ are dimensionless constants that depend on the lattice structure. 
The renormalized coupling constants are thus determined as  $\bar{g}=\beta g$ and $\bar{g}_{\ua\da}=\beta_{\ua\da}g_{\ua\da}$. 

%************************************************
\subsection{Elastic energy}\label{sec:elastic}
%************************************************

% [ Subsection introduction ]--------------------------------------
The expression of the elastic energy density $\Ecal_\elastic$ has been determined in Refs.\ \cite{Oktel06,Yoshino19}, and we summarize it in the following. 
% [ Expression of the elastic energy ]--------------------------------------
We first note that the elastic energy must be 
%Since the elastic energy is 
invariant under a constant change in $\uv_\alpha(\rv,t)$, i.e., translation of the lattices. 
Therefore, to the leading order in the derivative expansion, 
$\Ecal_\elastic$ should be a function of $\partial_i \uv_+~(i=x,y)$ and $\uv_-$, 
%It can therefore be decomposed as
resulting in the decomposition
\begin{equation}\label{eq: eff energy elastic}
 \Ecal_\elastic 
 =\Ecal_\elastic^{(+)}(\partial_i \uv_+)+\Ecal_\elastic^{(-)}(\uv_-)+\Ecal_\elastic^{(+-)}(\partial_i\uv_+,\uv_-). 
\end{equation}
To express $\Ecal_\elastic^{(+)}$, it is useful to introduce
\begin{equation}
 %w_0:=\partial_x u_+^x+\partial_y u_+^y,~~
 w_1:=\partial_x u_+^x -\partial_y u_+^y,~~
 w_2:=\partial_y u_+^x+\partial_x u_+^y. 
\end{equation}
%In the LLL regime, the vortex density stays constant, and therefore $w_0=0$; this can also be confirmed by using Eq.~\eqref{eq:EOM para/anti}. 
%From 
On the basis of
a symmetry consideration, each term in Eq.~\eqref{eq: eff energy elastic} can be expressed as
\begin{subequations}\label{eq : elastic energy}
\begin{align}
 &\Ecal_\elastic^{(+)} (\partial_i \uv_+) 
 = \frac{gn^2}{2} \left( C_1 w_1{}^2 + C_2 w_2{}^2 + C_3 w_1w_2 \right) , \\
 &\Ecal_\elastic^{(-)} (\uv_-) 
 = \frac{gn^2}{2\ell^2} \left[D_1 \left(u_-^x\right)^2+ D_2 \left(u_-^y\right)^2+D_3 u_-^xu_-^y\right], \\
 &\Ecal_\elastic^{(+-)} (\partial_i \uv_+,\uv_-) 
 = \frac{gn^2}{2\ell} F_1 \left(w_1u_-^y + w_2u_-^x \right).
\end{align}
\end{subequations}
For each of the vortex-lattice structures in Fig.\ \ref{fig:Phases_GP}(a)-(e), the dimensionless elastic constants $\{C_1, C_2, C_3, D_1, D_2,D_3,F_1\}$ satisfy 
\begin{equation}\label{eq:ElasEngy_ElasConst}
\begin{split}
 &\text{(a)}~C_1=C_2\equiv C>0,~D_1=D_2\equiv D>0,~C_3=D_3=F_1=0;\\
 &\text{(b)}~C_1=C_2\equiv C>0,  D_1=D_2\equiv D>0,~C_3=D_3=0,~F_1 \ne 0;\\
 &\text{(c)}~C_1, C_2, D_1,D_2>0,~ C_3,D_3\ne 0,~ F_1=0;\\
 &\text{(d)}~C_1, C_2>0,~ D_1=D_2\equiv D>0,~C_3=D_3=F_1=0;\\
 &\text{(e)}~C_1, C_2>0,~ D_1,D_2>0,~C_3=D_3=F_1=0.
\end{split}
\end{equation}

% [ Equivalence between the parallel- and antiparallel-field cases ]--------------------------------------
Ke\c{c}eli and Oktel \cite{Oktel06} have determined the constants $\{C_1, C_2, C_3, D_1, D_2,D_3\}$ 
by calculating a change in the mean-field energy under deformation of vortex lattices. 
In our previous work \cite{Yoshino19}, we have pointed out the presence of the $F_1$ term for interlaced triangular vortex lattices (b), 
which was missed in Ref.\ \cite{Oktel06}. 
For $\nu\gg 1$, the elastic constants should %take the same values between the parallel- and antiparallel-field cases 
be the same for the two types of fields 
because of the exact correspondence of the GP energy functionals \cite{Furukawa14}. %between the two cases 
In Sec.\ \ref{sec:numerics_excitation}, we will determine all the elastic constants above 
as a function of $g_{\ua\da}/g$ by using the data of the excitation spectra. 

%************************************************
\subsection{Diagonalization of the effective Hamiltonian}
%************************************************

% [ Fourier expansions ]--------------------------------------
We are now in a position to calculate the energy spectrum of the Hamiltonian $H=\int d^2\rv~{\cal H}$, 
where the Hamiltonian density ${\cal H}$ is given by Eq.\ \eqref{eq:Hcal_n_theta}.
We perform Fourier expansions
\begin{equation}
  \theta_\nu(\rv)=\frac{1}{\sqrt{A}}\sum_\kv \theta_{\kv,\nu}\erm^{\irm\kv\cdot\rv} , ~~
         n_\nu(\rv)=\frac{1}{\sqrt{A}}\sum_\kv        n_{\kv,\nu}\erm^{\irm\kv\cdot\rv} ~~ (\nu=\pm),
\end{equation}
where the Fourier components satisfy 
\begin{equation}
 [\theta_{\kv,\nu},n_{-\kv',\nu'}]=\irm \delta_{\nu\nu'}\delta_{\kv\kv'}, ~~
 \theta^\dagger_{\kv,\nu}=\theta_{-\kv,\nu},~~
        n^\dagger_{\kv,\nu}=       n_{-\kv,\nu} ~~ (\nu,\nu'=\pm) . 
\end{equation}
We note that the $\kv=\zerov$ component $n_{\zerov,\pm}$ of the densities is related to the atom numbers as $n_{\zerov,\pm}=(N_\ua\pm N_\da)/(2\sqrt{A})$. 
The Hamiltonian $H$ is then expressed as
\begin{equation}\label{eq:H_vorlat2}
\begin{split}
 H=&\frac{n}2\sum_{\kv}
   \begin{pmatrix}\theta_{-\kv,+}&\theta_{-\kv,-}\end{pmatrix}
   \begin{pmatrix}\Gamma_\pm(\kv)&\pm \irm\Gamma(\kv)\\
                           \mp \irm\Gamma(\kv)&\Gamma_\mp(\kv)\end{pmatrix}
   \begin{pmatrix}\theta_{ \kv,+} \\\theta_{ \kv,-}\end{pmatrix} 
 +\frac{1}{2n}\sum_{\kv} \sum_{\nu=\pm} e_{\kv,\nu} n_{-\kv,\nu} n_{\kv,\nu} ,
\end{split}
\end{equation}
where\footnote{We slightly change the definitions of $\Gamma_\pm(\kv)$ and $\Gamma(\kv)$ from Ref.\ \cite{Yoshino19}
by dividing them by $n$ so that they have the dimension of energy. }
\begin{subequations}\label{eq:Gamma}
\begin{align}
 e_{\kv,\nu}        &:=2\gb_\nu n+\frac{\hbar^2\kv^2}{2M},\\
 \Gamma_+(\kv)&:=gn\ell^4 
 \left[ C_1\left(2k_xk_y\right)^2+C_2\left(k_x^2-k_y^2\right)^2-C_3 (2k_xk_y) \left(k_x^2-k_y^2\right) \right],\\
 \Gamma_-(\kv)&:= gn\ell^2 \left( D_1k_y^2+ D_2k_x^2- D_3 k_xk_y \right),\\
 \Gamma   (\kv)&:= \frac12 gn\ell^3 F_1 \left[(2k_xk_y)k_x+(k_x^2-k_y^2)k_y \right] .
\end{align}
\end{subequations}
Here, the upper and lower signs correspond to the cases of parallel and antiparallel fields, respectively.\footnote{The same sign rule applies to 
Eqs.\ \eqref{eq:M_Gamma}, \eqref{eq:Gamma0z}, \eqref{eq:diagonalize_M_parameter}, \eqref{eq:m_k}, \eqref{eq:e_k_CD}, \eqref{eq:f12phi}, \eqref{eq:zeta_R_vorlat}, 
\eqref{eq:BoseDist._VorLat}, \eqref{eq:xi_F_G_vorlat}, \eqref{eq:cFG_CD}, \eqref{eq:Se_vorlat}, and \eqref{eq:dep_log} below.} 

% [ Zero-mode part ]--------------------------------------
It is useful to decompose the Hamiltonian \eqref{eq:H_vorlat2} into the zero-mode ($\kv=\zerov$) and oscillator-mode ($\kv\ne\zerov$) parts. 
First, the zero-mode part is given by 
\begin{equation}
 H^\zero= \sum_{\nu=\pm} \gb_\nu n_{\zerov,\nu}^2 =\frac{\gb_+}{4A}N^2+\frac{\gb_-}{4A}(N_\ua-N_\da)^2.
\end{equation}
Thus, the zero-mode energy is specified by the atom numbers $N_\ua$ and $N_\da$. 
In our setting of balanced population $N_\ua=N_\da$, 
the zero-mode state is given by the product state $\left|N_\ua=N/2\right\rangle  \left|N_\da=N/2\right\rangle$, 
which has no intercomponent entanglement. 

%In this case, both $N_\ua$ and $N_\da$ are conserved and thus the zero-mode ground state is the product state given simply by 
%\begin{equation}\label{eq:zero_mode}
% \ket{0^\zero}= \left|N_\ua=N/2\right\rangle  \left|N_\da=N/2\right\rangle .
%\end{equation}

% [ Oscillator-mode part ]--------------------------------------
Next, we discuss the oscillator-mode part $H^\osc$ of the Hamiltonian \eqref{eq:H_vorlat2}. 
To treat this part, we perform canonical transformations in two steps. 
The first transformation reads 
\begin{equation}\label{eq:CanoTrnf_Scaling}
 \tilde{\theta}_{\kv,+}=r_\kv^{-1} \theta_{\kv,+},~
 \tilde{\theta}_{\kv,- }=r_\kv        \theta_{\kv,-},~
 \tilde{       n}_{\kv,+}=r_\kv        n_{\kv,+} ,~
 \tilde{       n}_{\kv,- }=r_\kv^{-1} n_{\kv,-} ,
\end{equation}
where $r_\kv:=(e_{\kv,+}/e_{\kv,-})^{1/4}$. Then, $H^\osc$ is rewritten as 
\begin{equation}\label{eq:VorLat_Ham_2}
 H^\osc
 =\frac{n}{2}\sum_{\kv\neq\zerov}
 \begin{pmatrix}\tilde{\theta}_{-\kv,+}&\tilde{\theta}_{-\kv,-}\end{pmatrix} M(\kv)
 \begin{pmatrix}\tilde{\theta}_{ \kv,+}\\ \tilde{\theta}_{ \kv,-}\end{pmatrix} 
 +\frac{1}{2n}\sum_{\kv\neq\zerov}\sum_{\nu=\pm}e_\kv\tilde{n}_{-\kv,\nu}\tilde{n}_{\kv,\nu},
\end{equation}
where $e_\kv:=\sqrt{e_{\kv,+}e_{\kv,-}}$. Here, the $2\times2$ matrix $M(\kv)$ is given by
\begin{equation}\label{eq:M_Gamma}
 M(\kv)
 :=\begin{pmatrix} r_\kv^2 \Gamma_\pm(\kv)& \pm \irm\Gamma(\kv) \\
                              \mp \irm\Gamma(\kv)  & r_\kv^{-2} \Gamma_\mp(\kv)\end{pmatrix}
 =\Gamma_0(\kv) I 
   \mp \Gamma(\kv)\sigma_y
   + \Gamma_z(\kv) \sigma_z ,  
\end{equation}
where $I$ is the identity matrix, $(\sigma_x,\sigma_y,\sigma_z)$ are the Pauli matrices, and
\begin{equation}\label{eq:Gamma0z}
 \Gamma_0(\kv) := \frac12 \left[ r_\kv^2\Gamma_\pm(\kv)+r_\kv^{-2}\Gamma_\mp(\kv) \right],~~
 \Gamma_z(\kv) := \frac12 \left[ r_\kv^2\Gamma_\pm(\kv)-r_\kv^{-2}\Gamma_\mp(\kv) \right].
\end{equation}
We then perform the second canonical transformation using the unitary matrix $U(\kv)$ as
\begin{equation}\label{eq:CanoTrnf_Diagonalization}
 \begin{pmatrix} \tilde{\theta}_{\kv,+} \\ \tilde{\theta}_{\kv,-} \end{pmatrix} 
 = U(\kv)
 \begin{pmatrix} \bar{ \theta}_{\kv,1} \\ \bar{ \theta}_{\kv,2} \end{pmatrix} ,~~
 \begin{pmatrix} \tilde{n}_{\kv,+} \\ \tilde{n}_{\kv,-} \end{pmatrix} 
 = U(\kv)
 \begin{pmatrix} \bar{ n}_{\kv,1} \\ \bar{ n}_{\kv,2} \end{pmatrix}. 
\end{equation}
We note that the second term of Eq.\ \eqref{eq:VorLat_Ham_2} is invariant under this transformation if $U(-\kv)^t U(\kv)=I$ for all $\kv\ne\zerov$. 
It is therefore useful to choose $U(\kv)$ in such a way as to diagonalize the Hermitian matrix $M(\kv)$ as
\begin{equation}\label{eq:diagonalize_M}
 U^{-1}(\kv)M(\kv)U(\kv)
 =\begin{pmatrix} m_{\kv,1}& 0\\ 0& m_{\kv,2}\end{pmatrix} , ~~
 U(\kv)=\erm^{\irm\chi_\kv\sigma_x/2}
 =\begin{pmatrix} \cos(\chi_\kv/2)& \irm\sin(\chi_\kv/2)\\
                            \irm\sin(\chi_\kv/2)& \cos(\chi_\kv/2)\end{pmatrix} ,  
\end{equation}
where\footnote{
Since $\Gamma_z(\kv)=\Gamma_z(-\kv)$ and $\Gamma(-\kv)=-\Gamma(\kv)$, we find $\chi_{-\kv}=-\chi_{\kv}$ 
and thus $U(-\kv)^t=U(\kv)^\dagger=U(\kv)^{-1}$; therefore, the aforementioned condition $U(-\kv)^t U(\kv)=I$ is met. 
}
\begin{subequations}\label{eq:diagonalize_M_parameter}
\begin{align}
 &m_{\kv,j}
 := \Gamma_0(\kv) +(-1)^{j-1} \Lambda(\kv)~~(j=1,2),\\
 &\Lambda(\kv) (\cos\chi_\kv, \sin\chi_\kv) := \left( \Gamma_z(\kv),\Gamma(\kv) \right).
% \cos\chi_\kv =\frac{\Gamma_z(\kv)}{\Lambda(\kv)},~
% \sin \chi_\kv=\mp\frac{\Gamma(\kv)}{\Lambda(\kv)},~
% \Lambda(\kv)= \Gamma(\kv)^2+\Gamma_z(\kv)^2\right]^{1/2}  .
\end{align} 
\end{subequations}
In terms of the new set of canonical variables, the Hamiltonian \eqref{eq:VorLat_Ham_2} is further rewritten as 
\begin{equation}\label{eq:Hosc_thetab_nb}
 H^\osc
 =\frac12\sum_{\kv\neq\zerov}\sum_{j=1,2} 
 \left(n m_{\kv,j}\bar{\theta}_{-\kv,j}\bar{\theta}_{\kv,j}
     + \frac{e_\kv}{n}\bar{n}_{-\kv,j}\bar{n}_{\kv,j} \right) .
\end{equation}
Here, $\bar{\theta}_{\kv,j}$ and $\bar{n}_{\kv,j}$ satisfy 
$[\bar{\theta}_{\kv,j},\bar{n}_{-\kv',j'} ]= \irm \delta_{jj'}\delta_{\kv\kv'} ~(\kv,\kv'\ne\zerov;~j,j'=1,2)$ 
as the transformations in Eqs.\ \eqref{eq:CanoTrnf_Scaling} and \eqref{eq:CanoTrnf_Diagonalization} leave the commutation relations unchanged. 
Finally, introducing the (bogolon) annihilation and creation operators 
\begin{equation}\label{eq:CanoTrnf_VortexLatticeBogolon}
 \gamma_{\kv,j}
 :=\frac{1}{\sqrt{2}}\left(\sqrt{n}\zeta_{\kv,j}\bar{\theta}_{\kv,j} 
   +\frac{\irm\bar{n}_{ \kv,j}}{\sqrt{n}\zeta_{\kv,j}}\right) , ~
 \gamma_{\kv,j}^\dagger
 :=\frac{1}{\sqrt{2}}\left(\sqrt{n}\zeta_{\kv,j}\bar{\theta}_{-\kv,j} 
   - \frac{\irm\bar{n}_{-\kv,j}}{\sqrt{n}\zeta_{\kv,j}}\right) ~(\kv\ne\zerov;~j=1,2),
\end{equation}
with $\zeta_{\kv,j}:=(m_{\kv,j}/e_\kv)^{1/4}$, 
we can diagonalize the Hamiltonian \eqref{eq:Hosc_thetab_nb} as
\begin{equation}\label{eq:Hosc_gamma}
 H^\osc
 =\sum_{\kv\neq\zerov}\sum_{j=1,2}
 E_j(\kv) \left(\gamma_{\kv,j}^\dagger\gamma_{\kv,j}+\frac12\right) , ~~
 E_j(\kv):=\sqrt{m_{\kv,j}e_\kv} . 
\end{equation}
The ground state $\ket{0^\osc}$ of this Hamiltonian is specified by the condition that $\gamma_{\kv,\pm}\ket{0^\osc}=0$ for all $\kv\ne\zerov$. 

% [ Energy spectrum ]--------------------------------------
We now discuss the single-particle spectrum $E_j (\kv)~(j=1,2)$ in the long-wavelength limit $k\ell \ll 1$.  
In this limit, we can make the approximation $r_\kv\approx r_\zerov=\left(\gb_+/\gb_-\right)^{1/4}$ and $e_\kv\approx e_\zerov = 2(\gb_+\gb_-)^{1/2}n$.
%$\tilde{\Gamma}_+(\kv)/\tilde{\Gamma}_-(\kv)={\cal O}(k^2)$ and  $\Gamma(\kv)/\tilde{\Gamma}_-(\kv)={\cal O}(k)$
Since $\Gamma_+(\kv)=\Ocal(k^4)$, $\Gamma_-(\kv)=\Ocal(k^2)$, and $\Gamma(\kv)=\Ocal(k^3)$ as seen in Eq.\ \eqref{eq:Gamma}, 
we can approximate $m_{\kv,j}$ in Eq.\ \eqref{eq:diagonalize_M_parameter} as
\begin{align}\label{eq:m_Gamma_k}
 m_{\kv,1}\approx r_\zerov^{\mp 2} \Gamma_-(\kv),~~
 m_{\kv,2}\approx r_\zerov^{\pm 2} \left[ \Gamma_+(\kv) -\frac{\Gamma(\kv)^2}{\Gamma_-(\kv)} \right].
\end{align}
By parametrizing the wave vector as $(k_x,k_y)=k(\cos\varphi,\sin\varphi)$, $m_{\kv,j}$ can also be expressed as
\begin{equation}\label{eq:m_k}
 m_{\kv,1}\approx  r_\zerov^{\mp 2} gn D(\varphi) (k\ell)^2,~~
 m_{\kv,2}\approx  r_\zerov^{\pm 2} gn C(\varphi) (k\ell)^4, 
\end{equation}
where 
\begin{subequations}\label{eq:CD_phi}
\begin{align}
 C(\varphi)&:=C_1\sin^2(2\varphi)+C_2\cos^2(2\varphi)
  -C_3\sin(2\varphi)\cos(2\varphi)-C_4\sin^2(3\varphi) , \\
 D(\varphi)&:=D_1\sin^2(\varphi)+D_2\cos^2(\varphi)-D_3\sin(\varphi)\cos(\varphi) , 
\end{align}
\end{subequations} 
with $C_4:=F_1^2/4D_1$. We then obtain the low-energy %dispersion relations 
spectrum 
as 
\begin{equation}\label{eq:e_k_CD}
 E_j(\kv) \approx\sqrt{m_{\kv,j} e_\zerov}\approx \sqrt2 gn(k\ell)^j f_j(\varphi)~~~(j=1,2) ,
\end{equation}
where the dependence on the angle $\varphi$ is expressed by the dimensionless functions 
\begin{equation}\label{eq:f12phi}
 f_1(\varphi)=\sqrt{\gb_\mp D(\varphi)/g}, ~~~f_2(\varphi)=\sqrt{\gb_\pm C(\varphi)/g}. 
\end{equation}
We thus find that low-energy modes with linear and quadratic dispersion relations emerge with anisotropy that depends on the lattice structure. 
Furthermore, the low-energy dispersion relations for parallel (P) and antiparallel (AP) fields are related by proper rescaling as follows: 
\begin{equation}\label{eq:E12_rescaling}
 E_1^\mathrm{P}(\kv)/\sqrt{\gb_-}=E_1^\mathrm{AP}(\kv)/\sqrt{\gb_+},~~~
 E_2^\mathrm{P}(\kv)/\sqrt{\gb_+}=E_2^\mathrm{AP}(\kv)/\sqrt{\gb_-}.
\end{equation}
Using the dimensionless functions $f_j(\varphi)~(j=1,2)$ for the two types of fields, these relations can also be written as
\begin{equation}\label{eq:energy_rescaling}
\begin{split}
 f_1^\mathrm{P}(\varphi)\sqrt{\frac{g}{\gb_-}} =f_1^\mathrm{AP}(\varphi)\sqrt{\frac{g}{\gb_+}}=\sqrt{D(\varphi)},~~~
 f_2^\mathrm{P}(\varphi)\sqrt{\frac{g}{\gb_+}}=f_2^\mathrm{AP}(\varphi)\sqrt{\frac{g}{\gb_-}} =\sqrt{C(\varphi)}.
\end{split}
\end{equation}
In Sec.\ \ref{sec:numerics_excitation},  we will confirm these relations through the numerical calculations 
by the Bogoliubov theory for all the five vortex-lattice structures in Fig.\ \ref{fig:Phases_GP}(a)-(e). 
While similar rescaling relations are also discussed in Ref.\ \cite{Yoshino19}, 
the importance of using the renormalized coupling constants $\gb_\pm$ is overlooked there. 
Without the renormalization, the rescaling relations in Eqs.\ \eqref{eq:E12_rescaling} and \eqref{eq:energy_rescaling} are satisfied 
only for overlapping vortex lattices where $\beta=\beta_{\ua\da}$, 
as confirmed numerically in Ref.\ \cite{Yoshino19}. 

%The angular part of the energy spectrum by BdG theory have been rescaled by sqrt{2} in the figure.
%\begin{equation}%\label{eq:e_k_CD}
% E_j(\kv)\approx\sqrt{2(g_+g_-)^{1/2}m_{\kv,j}}\approx  
% \begin{cases}
%  [2gg_\mp D(\varphi)]^{1/2} nk\ell       & (j=1) ; \\
% [2gg_\pm C(\varphi)]^{1/2} n(k\ell)^2 & (j=2) , \\
% \end{cases}
%\end{equation}

%%%%%%%%%%%%%%%%%%%%%%%%%%%%%%%%%%%%%%%%%%%%%%%%
\subsection{Intercomponent entanglement}\label{sec:ent_vorlat}
%%%%%%%%%%%%%%%%%%%%%%%%%%%%%%%%%%%%%%%%%%%%%%%%

% [ Reduced density matrix ]--------------------------------------
We now calculate the reduced density matrix (RDM) $\rho_\ua$ for the spin-$\ua$ component, 
which is defined by starting from the ground state $\ket{0^\zero}\otimes\ket{0^\osc}$ of the total system and trancing out the degrees of freedom in the spin-$\da$ component. 
We then discuss the properties of the intercomponent entanglement. 
Because of the decoupling of the zero and oscillator modes, 
the RDM takes the form of $\rho_\ua = \rho_\ua^{\mathrm{zero}} \otimes \rho_\ua^{\mathrm{osc}}$. 
As the zero-mode ground state $\left|N_\ua=N/2\right\rangle  \left|N_\da=N/2\right\rangle$ is a product state, there is no intercomponent entanglement in the zero-mode part; 
the RDM in this part is given simply by  $\rho_\ua^\zero=\ket{N_\ua=N/2}\bra{N_\ua=N/2}$. 
Below we consider the oscillator-mode part $\rho_\ua^{\mathrm{osc}}$. 

%First, the zero-mode part has no contribution to the entanglement Hamiltonian as the zero-mode ground state is given by the product state \eqref{eq:zero_mode}. 

% [ Ansatz for the RDM ]--------------------------------------
For $\rho_\ua^\osc$, we introduce the following Gaussian ansatz \cite{Yoshino21,Chen13,Lundgren13,Peschel03,Peschel09,Metlitski11}: 
\begin{equation}\label{eq:Entanglement_Ham_oscillation}
 \rho^{\osc}_\ua =\frac1{Z^{\osc}_\mathrm{e}} \mathrm{e}^{-H_\mathrm{e}^\osc} , ~Z^\osc_\mathrm{e}=\Tr\mathrm{e}^{-H^{\osc}_\mathrm{e}} ,~
 H^{\mathrm{osc}}_\mathrm{e}
 =\frac12\sum_{\kv\ne\zerov}
 \left(nF_\kv\theta_{-\kv,\ua}\theta_{\kv,\ua}+\frac{G_\kv}{n}n_{-\kv,\ua}n_{\kv,\ua}\right) ,
\end{equation} 
where $F_\kv$ and $G_\kv$ are positive dimensionless coefficients to be determined later 
and we assume $F_\kv=F_{-\kv}$ and $G_\kv=G_{-\kv}$ for convenience. 
%(we assume $F_\kv=F_{-\kv}$ and $G_\kv=G_{-\kv}$ without loss of generality). 
By introducing annihilation and creation operators as 
\begin{subequations}\label{eq:eta_k}
\begin{align}
  \eta_\kv=&\frac{1}{\sqrt{2}} \left[
    \sqrt{n}             \left(\frac{F_\kv}{G_\kv}\right)^{1/4} \theta_{\kv,\ua} +
    \frac{\irm}{\sqrt{n}}\left(\frac{G_\kv}{F_\kv}\right)^{1/4} n_{\kv,\ua} \right], \\
  \eta_\kv^\dagger=&\frac{1}{\sqrt{2}} \left[
    \sqrt{n}             \left(\frac{F_\kv}{G_\kv}\right)^{1/4} \theta_{-\kv,\ua} -
    \frac{\irm}{\sqrt{n}}\left(\frac{G_\kv}{F_\kv}\right)^{1/4} n_{-\kv,\ua} \right] ~~~(\kv\ne\zerov), 
\end{align}
\end{subequations}
the entanglement Hamiltonian $H_\erm^\osc$ in Eq.\ \eqref{eq:Entanglement_Ham_oscillation} is diagonalized as   
\begin{equation}\label{eq:DiagonalHam}
 H^{\osc}_\erm=\sum_{\kv\ne\zerov} \xi_\kv \left(\eta_\kv^\dagger\eta_\kv+\frac12\right) ,
\end{equation}
where $\xi_\kv:=\sqrt{F_\kv G_\kv}$ is the single-particle ES. 

The single-particle ES $\xi_\kv$ and the coefficients $F_\kv$ and $G_\kv$ can be determined in the following way \cite{Yoshino21}. 
Using the relations in Eq. \eqref{eq:eta_k} and the Bose distribution function 
\begin{equation}\label{eq:fB_xi}
 \Tr\left(\eta_\kv^\dagger \eta_\kv \rho^\osc_\ua\right)=\frac{1}{\erm^{\xi_\kv}-1} \equiv f_\Brm (\xi_\kv) ,
\end{equation}
we obtain the phase and density correlators as
\begin{subequations}\label{eq:Correlator_FkGk}
\begin{align}
 &\Tr\left(\theta_{-\kv,\ua}\theta_{ \kv,\ua}\rho^\osc_\ua\right)
 %=\frac{1}{2n}\left(\frac{G_\kv}{F_\kv}\right)^{1/2} \!\!
 %  \Tr\left[\left(\eta_{ \kv}^\dagger+\eta_{-\kv}\right) 
 %             \left(\eta_{ \kv}+\eta_{-\kv}^\dagger\right)\rho^\osc_\ua\right] 
 =\frac{1}{n} \left(\frac{G_\kv}{F_\kv}\right)^{1/2} \! \left[f_\Brm(\xi_\kv)+\frac12\right],\\
 &\Tr\left(n_{-\kv,\ua}n_{\kv,\ua}\rho^\osc_\ua\right)
 %=\frac{n}{2}\left(\frac{F_\kv}{G_\kv}\right)^{1/2} \!\!
 %    \Tr\left[\left(\eta_{ \kv}^\dagger-\eta_{-\kv} \right)
 %                \left(\eta_{ \kv} -\eta_{-\kv}^\dagger\right) 
 %               \rho^\osc_\ua\right]  
 =n\left(\frac{F_\kv}{G_\kv}\right)^{1/2} \! \left[f_\Brm(\xi_\kv)+\frac12\right] . 
\end{align} 
\end{subequations}
We can then determine $f_\Brm(\xi_\kv)$ and $F_\kv / G_\kv$ by requiring these correlators to be equal 
to the same correlators calculated for the oscillator ground state $\ket{0^\mathrm{osc}}$ of the total system. 
Details of this calculation are described in \ref{app:entanglement}. 
In the long-wavelength limit $k\ell\ll 1$, we obtain 
\begin{equation}\label{eq:xi_F_G_vorlat}
 \xi_\kv \approx  c(\varphi) \sqrt{k\ell}, ~~
 F_\kv\approx  F(\varphi) (k\ell)^2 ,~~
 G_\kv \approx  \frac{G(\varphi)}{k\ell}, 
\end{equation}
where the dependences on the angle $\varphi$ are expressed by the dimensionless functions
\begin{equation}\label{eq:cFG_CD}
 c(\varphi)=4\left[\frac{\gb_\mp C(\varphi)}{\gb_\pm D(\varphi)}\right]^{1/4},~
 F(\varphi)=4\sqrt{ \frac{2gC(\varphi)}{\gb_\pm} },~
 G(\varphi)=2\sqrt{ \frac{2\gb_\mp}{gD(\varphi)} }.
\end{equation}
We find that the ES shows a gapless square-root dispersion relation with anisotropy that depends on the lattice structure. 
Furthermore, similarly to the case of excitation spectra [see Eq.\ \eqref{eq:E12_rescaling}],
the single-particle ES $\xi_\kv$ for parallel (P) and antiparallel (AP) fields are related by suitable rescaling as
\begin{equation}
 \left( \frac{\gb_+}{\gb_-} \right)^{1/4} \xi_\kv^\mathrm{P} = \left( \frac{\gb_-}{\gb_+} \right)^{1/4} \xi_\kv^\mathrm{AP} .
\end{equation}
Using the dimensionless functions $c(\varphi)$ for the two types of fields, this relation can also be written as
\begin{equation}\label{eq:c_phi_rescale}
     \left(\frac{\gb_+}{\gb_-}\right)^{1/4} c^\mathrm{P}(\varphi)
 =\left(\frac{\gb_-}{\gb_+}\right)^{1/4} c^\mathrm{AP}(\varphi)
 = 4\left[\frac{C(\varphi)}{D(\varphi)}\right]^{1/4}. 
\end{equation}

% [ Entanglement Hamiltonian ]--------------------------------------
The entanglement Hamiltonian is given in the long-wavelength limit 
by $H_\erm^\osc$ in Eq.\ \eqref{eq:Entanglement_Ham_oscillation} with $F_\kv$ and $G_\kv$ in Eq.\ \eqref{eq:xi_F_G_vorlat}. 
Using the fields $\theta_\ua(\rv)$ and $n_\ua(\rv)$ in real space, it can be expressed as
\begin{equation}\label{eq:He_vorlat}
\begin{split}
 H_\erm
 =\int\mathrm{d}^2\rv \int\mathrm{d}^2\rv'
   \Biggl[\frac{n\ell^2}{2} U_F(\rv-\rv') \nabla\theta_\ua(\rv)\cdot\nabla\theta_\ua(\rv') 
  +\frac{1}{2n} U_G(\rv-\rv') n_\ua(\rv)n_\ua(\rv')   \Biggr] , 
\end{split}
\end{equation}
where we introduce the interaction potentials 
\begin{equation}\label{eq:VorLat_Potential}
\begin{split}
 U_F(\rv-\rv') =\frac{1}{A} \sum_{\kv} F(\varphi) \erm^{\irm\kv\cdot(\rv-\rv')} , ~~
  %= \frac{\ell^2}{(2\pi)^2}\int d^2\kv\ \sqrt{C(\varphi)}e^{i\kv(\rv'-\rv)} , \\
 U_G(\rv-\rv') =\lim_{\alpha\to 0^+} \frac1{A} \sum_{\kv\neq\zerov} \frac{G(\varphi)}{k\ell}
  \erm^{-\alpha k+\irm\kv\cdot(\rv-\rv')}.
  %=\lim_{\alpha\to0} \frac{1}{(2\pi)^2\ell}\int d^2\kv\ \frac{e^{-\alpha|\kv|+i\kv(\rv'-\rv)}}{k\sqrt{D(\varphi)}}  .
\end{split}
\end{equation}
Here, we use the convergence factor $\erm^{-\alpha k}$ to regularize the infinite sum for $U_G(\rv-\rv')$. 
For simplicity, we consider the case of overlapping triangular lattices, in which $C(\varphi)$ and $D(\varphi)$ [and thus $F(\varphi)$ and $G(\varphi)$ as well] are constant; see (a) in Eq.\ \eqref{eq:ElasEngy_ElasConst}. 
In this case, the potentials in Eq.\ \eqref{eq:VorLat_Potential} are calculated as 
\begin{equation}\label{eq:VorLat_Potential2}
  U_F(\rv-\rv')=F \delta(\rv-\rv') , ~~U_G(\rv-\rv')=\frac{G}{2\pi\ell |\rv-\rv'|} .
\end{equation}
For the calculation of $U_G(\rv-\rv')$, we refer the reader to Appendix A of Ref.\ \cite{Yoshino21}. 
We note that $\theta_\ua(\rv)$ is the regular part of the superfluid phase of the spin-$\ua$ component, 
and that its gradient is related to the regular part of the superfluid velocity, $\vv_{\srm,\ua}(\rv)=-\frac{\hbar}{M}\nabla\theta_\ua(\rv)$. 
Therefore, the entanglement Hamiltonian \eqref{eq:He_vorlat} has a short-range interaction in terms of the superfluid velocity $\vv_{s,\ua}(\rv)$ 
and a long-range one in terms of the density $n_\ua(\rv)$. 
If the density interaction were short-ranged, the ES would show a phonon mode with a linear dispersion relation. 
Therefore, the anomalous square-root dispersion relation in Eq.\ \eqref{eq:xi_F_G_vorlat} is closely related with the presence of a long-range interaction in $H_\erm$. 

%The presence of a long-range interaction is crucial for the emergence of the anomalous square-root dispersion relation in Eq.\ \eqref{eq:xi_F_G_vorlat}. 

% # We can not explicitly compare the H_ent with the original Hamiltonian.
% # How do we emphasize the relation of the long-range interaction with the square-root dispersion relation?

% [ Entanglement entropy ]--------------------------------------
Using the single-particle ES $\xi_\kv$ in Eq.\ \eqref{eq:xi_F_G_vorlat}, we can calculate the intercomponent EE $S_\erm$. 
For simplicity, we assume that $\xi_\kv$ is isotropic, i.e., $c(\varphi)$ is constant, as in the case of overlapping triangular lattices; 
however, we expect that the result holds qualitatively for all the lattice structures. 
With this simplification, the EE is calculated as  (see Appendix B of Ref.\ \cite{Yoshino21} for the derivation)
\begin{equation}\label{eq:Se_Nv_logNv}
 S_\erm = \frac{\sigma A}{c^4\ell^2} - \frac12 \ln \frac{\sqrt{A}}{2\pi c^2 \ell} + O(1) = \frac{2\pi\sigma \Nvor}{c^4} - \frac14 \ln \frac{\Nvor}{2\pi c^4} + O(1) . 
\end{equation}
Here, the leading contribution is given by the first term, which is proportional to the area $A$ with a non-universal coefficient $\sigma$ that depends, e.g., on the choice of the high-momentum cutoff. 
Besides, there is a subleading logarithmic term with the universal coefficient (equal to $-1/2$ when written as a function of the linear system size $\sqrt{A}$), 
which is identified through a careful examination of small-$k$ contributions and therefore originates from the Nambu-Goldstone modes. 
The intercomponent EE per flux in the thermodynamic limit is calculated as
\begin{equation}\label{eq:Se_vorlat}
 \lim_{\Nvor\to\infty} \frac{S_\erm}{\Nvor} = \frac{2\pi \sigma}{c^4} =\frac{\pi \sigma \gb_\pm D}{128 \gb_\mp C}  .
\end{equation}
Because of the factor $\gb_\pm/\gb_\mp$ in Eq.\ \eqref{eq:Se_vorlat}, when the intercomponent interaction $g_{\ua\da}$ is repulsive (attractive), 
the intercomponent EE is expected to be larger for the case of parallel (antiparallel) fields. 
We note that the above calculation of the intercomponent EE $S_\erm$ is simple yet approximate 
as it is based on the single-particle ES $\xi_\kv$ for long wavelengths. 
In Sec.\ \ref{sec:numerics_ent}, we will present numerical results on $S_\erm$ based on the Bogoliubov theory, 
in which $\xi_\kv$ over the full Brillouin zone is taken into account, 
and confirm the consistency with the field-theoretical predictions. 
%This behavior is in qualitative agreement with the numerical results in the quantum Hall regime \cite{Furukawa14,Furukawa17}. 

%%%%%%%%%%%%%%%%%%%%%%%%%%%%%%%%%%%%%%%%%%%%%%%%
\subsection{Intracomponent correlation functions}\label{sec:EFT_vorlat_corr}
%%%%%%%%%%%%%%%%%%%%%%%%%%%%%%%%%%%%%%%%%%%%%%%%

Here we calculate some intracomponent correlation functions, and discuss their connections with the (long-wavelength) entanglement Hamiltonian $H_\erm$ obtained in the preceding section. 
Let $\langle{\cal O}\rangle$ denote the expectation value of an operator ${\cal O}$ with respect to the ground state $\ket{0^\zero}\otimes\ket{0^\osc}$ of the total system. 
If ${\cal O}$ acts only on the spin-$\ua$ component, $\langle{\cal O}\rangle$ should be equal to $\Tr \left({\cal O} \erm^{-H_\erm}\right) / \Tr \erm^{-H_\erm}$ 
as far as long-distance properties are concerned.
Our purpose here is to investigate how the unusual long-range interactions in $H_\erm$ manifest themselves in the correlation properties of the system. 

% [ Correlations in k space ]--------------------------------------
Owing to the gapless ES $\xi_\kv$, we can approximate the Bose distribution function \eqref{eq:fB_xi} as 
$f_\Brm(\xi_\kv)\approx \xi_\kv^{-1}=(F_\kv G_\kv)^{-1/2}$ for sufficiently small $k$. 
Then, in the long-wavelength limit, Eq.\ \eqref{eq:Correlator_FkGk} gives 
\begin{align}\label{eq:corr_theta_n_vorlat_FG}
 \langle \theta_{-\kv,\ua}\theta_{ \kv,\ua} \rangle \approx \frac{1}{nF_\kv}\approx \frac{1}{nF(\varphi)k^2\ell^2},~~~
 \langle n_{-\kv,\ua} n_{\kv,\ua} \rangle \approx  \frac{n}{G_\kv}\approx \frac{nk\ell}{G(\varphi)}~~(\kv\ne\zerov), 
\end{align}
where we use Eq.\ \eqref{eq:xi_F_G_vorlat}. 
Therefore, the phase and density fluctuations are directly related to the coefficients $F_\kv$ and $G_\kv$, respectively, 
in the entanglement Hamiltonian \eqref{eq:Entanglement_Ham_oscillation}. 
Equation \eqref{eq:corr_theta_n_vorlat_FG} indicates that in the long-wavelength limit $k\to 0$, 
the phase fluctuation diverges and the density fluctuation is suppressed. 
From the viewpoint of the entanglement Hamiltonian, suppression of the density fluctuation is a consequence of the long-range interaction in terms of the density.  
The enhanced phase fluctuation in the long-wavelength limit leads to a quasi-long-range order in the one-particle density matrix, as we explain in the following. 

% [ One-particle density matrix ]--------------------------------------
The one-particle density matrix plays a key role in the characterization of the Bose-Einstein condensation \cite{Pethick_Smith_2008, Penrose56,Yang62}. 
To analyze its behavior in our course-grained description, we introduce the modified bosonic field $\tilde{\psi}_\ua=e^{-i\theta_\ua} \sqrt{n_\ua}$. 
As $\theta_\ua(\rv)$ is the regular part of the superfluid phase and $n_\ua(\rv)$ is the course-grained density, $\psit(\rv)$ is expected to vary slowly over space. 
We consider the modified one-particle density matrix
\begin{equation}\label{eq:corr_psit_theta_n}
 \langle \psit_\ua (\rv)^\dagger \psit_\ua (\zerov)\rangle=\left\langle \sqrt{n_\ua(\rv)} \erm^{i(\theta_\ua(\rv)-\theta_\ua(\zerov))} \sqrt{n_\ua(\zerov)}\right\rangle,
\end{equation}
which describes the slowly varying component of the ordinary one-particle density matrix. 
Its long-distance behavior is determined dominantly by the phase fluctuation 
as seen in the small-$k$ behavior of Eq.\ \eqref{eq:corr_theta_n_vorlat_FG}. 
Here we again consider the case of overlapping triangular lattices where $F(\varphi)$ is constant. 
By using Eq.\ \eqref{eq:corr_theta_n_vorlat_FG}, the phase correlation function in real space is obtained as
(see \ref{app:phase_corr} for the derivation) 
\begin{equation}\label{eq:corr_theta_r0_vorlat}
 \left\langle \left[ \theta_\ua(\rv)-\theta_\ua(\zerov) \right]^2\right\rangle 
 = \frac2A \sum_{\kv\ne\zerov} \erm^{-\alpha k} 
 \left[1- \cos \left({\kv\cdot\rv}\right) \right] \langle \theta_{-\kv,\ua}\theta_{ \kv,\ua} \rangle
 \approx \frac{1}{\pi nF\ell^2} \ln \frac{r}{2\alpha}~~(r\gg \alpha),
\end{equation}
where we again introduce the convergence factor $\erm^{-\alpha k}$ to regularize the infinite sum. 
The modified one-particle density matrix is then obtained as 
\begin{equation}\label{eq:corr_psi_vorlat}
 \langle \psit_\ua (\rv)^\dagger \psit_\ua (\zerov)\rangle 
 \approx n \exp \left\{ -\frac12 \left\langle \left[ \theta_\ua(\rv)-\theta_\ua(\zerov) \right]^2\right\rangle \right\}
 \approx n \left(\frac{r}{2\alpha} \right)^{-\frac{1}{2\pi nF\ell^2}}.
\end{equation}
We thus have a quasi-long-range order in the one-particle density matrix. 

% [ Depletion ]--------------------------------------
For a finite system of area $A$, the particle density $n_0$ of the condensate can be evaluated 
from the one-particle density matrix \eqref{eq:corr_psi_vorlat} at the large separation $r=\sqrt{A}/2$.
The density $n'$ of the depletion is then given by $n'=n-n_0$. 
Assuming $n'\ll n$ as required for the Bogoliubov theory and the present effective theory, 
the fraction of depletion is estimated as
\begin{equation}\label{eq:dep_log}
 \frac{n'}{n}
 \approx \frac12 \left\langle \left[ \theta_\ua(\rv)-\theta_\ua(\zerov) \right]^2\right\rangle \bigg|_{r=\sqrt{A}/2}
 \approx \frac{1}{2\pi nF\ell^2} \ln \frac{\sqrt{A}}{4\alpha} 
 = \frac{1}{2\nu} \left(\frac{\gb_\pm}{2gC}\right)^{1/2} \ln \frac{\sqrt{A}}{4\alpha} .
\end{equation}
where we use Eq.\ \eqref{eq:cFG_CD} and $\nu=2nA/\Nvor=4\pi n\ell^2$. 
For fixed $\nu$, the fraction of quantum depletion of the condensate increases logarithmically as a function of the vortex number $\Nvor=A/(2\pi\ell^2)$. 
Furthermore, for an intercomponent repulsion (attraction), it is larger for the case of parallel (antiparallel) fields, indicating larger quantum fluctuations. 
This behavior is in accord with the larger intercomponent EE given in Eq.\ \eqref{eq:Se_vorlat}. 
We note that a quasi-long-range order in the one-particle density matrix 
and a logarithmic increase of the fraction of depletion as in Eqs.\ \eqref{eq:corr_psi_vorlat} and \eqref{eq:dep_log} 
have also been discussed for a vortex lattice in a scalar BEC \cite{Sinova02,Baym04,Matveenko11,Kwasigroch12}. 

%************************************************
%\subsection{Comparison with homogeneous binary BECs}
%************************************************

% [ Comparison ]--------------------------------------
%It is interesting to compare the present results with the entanglement properties of binary BECs in the absence of synthetic gauge fields studied in Ref.\ \cite{Yoshino21}. 
%(i) In the presence of an intercomponent tunneling (a Rabi coupling), the ES has been found to show a square-root dispersion relation; 
%the entanglement Hamiltonian shows long- and short-range interactions in terms of the superfluid velocity and the density, respectively. 
%(ii) When the two components are coupled only by density-density interactions, the ES shows a gapped dispersion relation, 
%and the entanglement Hamiltonian has long-range interactions in terms of both the superfluid velocity and the density. 
%The difference of the present results from (ii) can be understood from the enhanced phase fluctuation due to the quadratic excitation mode. 

%%%%%%%%%%%%%%%%%%%%%%%%%%%%%%%%%%%%%%%%%%%%%%%%%
\section{Bogoliubov theory}\label{sec:BogoliubovTheory}
%%%%%%%%%%%%%%%%%%%%%%%%%%%%%%%%%%%%%%%%%%%%%%%%%

The Bogoliubov theory with the LLL approximation has been formulated for a scalar BEC in Refs.\ \cite{Sinova02, Matveenko11, Kwasigroch12}. 
In Ref.\ \cite{Yoshino19}, we have applied this theory to the present problem of binary BECs in parallel and antiparallel magnetic fields. 
In Secs.\ \ref{sec:LLL} and \ref{sec:BogoliubovTheory_spec}, we summarize the formulation of Ref.\ \cite{Yoshino19}. 
In particular, we give an expression of the ground-state energy [Eq.\ \eqref{eq:Egs/atom}]
in which a quantum correction due to zero-point fluctuations is included. 
In Sec.\ \ref{sec:BogoliubovTheory_ent}, 
we use this formulation to derive expressions of the intercomponent ES and EE. 
Numerical results based on this formulation will be presented in Sec.\ \ref{sec:NumericalResults}. 

%************************************************
\subsection{LLL magnetic Bloch states and Hamiltonian}\label{sec:LLL}
%************************************************

% [ LLL magnetic Bloch state ]--------------------------------------
We employ the LLL magnetic Bloch states $\{\Psi_{\kv\alpha}(\rv)\}$ \cite{Kwasigroch12,Rashba97,Burkov10,Panfilov16} 
as a convenient single-particle basis for describing vortex lattices. 
The expressions of these states are shown in Ref.\ \cite{Yoshino19}, and we summarize their main features in the following. 
Let $\mathbf{a}_1$ and $\mathbf{a}_2$ be the primitive vectors of a vortex lattice, 
and let $\Kv_\alpha$ be the pseudomomentum operator for a spin-$\alpha$ atom in a synthetic magnetic field $B_\alpha~(\alpha=\ua,\da)$. 
As expected for a ``Bloch state'', $\Psi_{\kv\alpha}(\rv)$ is an eigenstate of 
the magnetic translation $\erm^{-\irm \Kv_\alpha \cdot\av_j/\hbar}$ with an eigenvalue $\erm^{-i\kv\cdot\av_j}~(j=1,2)$. 
By taking $\Nvor$ discrete wave vectors $\kv$ consistent with the boundary conditions of the system, 
$\{\Psi_{\kv\alpha}(\rv)\}$ form a complete orthogonal basis of 
the LLL manifold.\footnote{
In numerical calculations presented later, we set $\kv=\frac{n_1}{\Nvoro} \bv_1+\frac{n_2}{\Nvort}\bv_2$ 
with $n_j\in\{0,1,\dots,N_{\mathrm{v}j}-1\}$ and $\Nvoro\Nvort=\Nvor$. 
Here, $\bv_1$ and $\bv_2$ are the reciprocal primitive vectors as shown in Fig.\ \ref{fig:Phases_GP}. 
}   
Notably, $\Psi_{\kv\alpha}(\rv)$ has a periodic pattern of zeros at \cite{Burkov10}
\begin{equation}\label{eq:zeros_Bloch}
 \rv=n_1\mathbf{a}_1+n_2\mathbf{a}_2+\frac12 (\av_1+\av_2) - \epsilon_\alpha \ell^2 \ev_z\times\kv,~~n_1,n_2\in\Zbb.
\end{equation} 
Therefore, $\Psi_{\kv\alpha}(\rv)$ represents a vortex lattice with primitive vectors $\av_1$ and $\av_2$ for any $\kv$, 
and the locations of vortices (zeros) can be shifted by varying $\kv$. 
Vortex lattices of binary BECs in Fig.\ \ref{fig:Phases_GP} are obtained 
when spin-$\alpha$ bosons condense into $\Psi_{\qv_\alpha,\alpha}(\rv)$, 
where the wave vectors $\qv_\ua$ and $\qv_\da$ are chosen in a way consistent 
with the displacement $u_1\av_1+u_2\av_2$ between the components. 

% [ Interaction Hamiltonian ]--------------------------------------
In the LLL approximation, the kinetic energy of each particle stays constant, 
and therefore we can focus on the interaction Hamiltonian $H_\mathrm{int}$. 
Using the LLL magnetic Bloch states $\{\Psi_{\kv\alpha}(\rv)\}$ as the basis, it is represented as
\begin{equation}
H_{\mathrm{int} } 
  =\frac{1}{2} \sum_{\alpha,\beta} \sum_{\mathbf{k}_1,\mathbf{k}_2,\mathbf{k}_3,\mathbf{k}_4} 
   V_{\alpha\beta}(\mathbf{k}_1,\mathbf{k}_2,\mathbf{k}_3,\mathbf{k}_4) 
   b^{\dag}_{\mathbf{k}_1 \alpha}b^{\dag}_{\mathbf{k}_2 \beta}b_{\mathbf{k}_3 \beta}b_{\mathbf{k}_4 \alpha} ,
\label{2ndQ hamiltonian}
\end{equation}
where $b_{\kv\alpha}$ is a bosonic annihilation operator for the state $\Psi_{\kv\alpha}(\rv)$ and 
\begin{equation}\label{V alpha beta}
  V_{\alpha\beta}(\mathbf{k}_1,\mathbf{k}_2,\mathbf{k}_3,\mathbf{k}_4)=
  g_{\alpha\beta} \int \drm^2\mathbf{r}~ 
  \Psi^{\ast}_{\mathbf{k}_1\alpha}(\mathbf{r}) \Psi^{\ast}_{\mathbf{k}_2\beta}(\mathbf{r}) 
  \Psi_{\mathbf{k}_3\beta}(\mathbf{r}) \Psi_{\mathbf{k}_4\alpha}(\mathbf{r}).
\end{equation}
The expression of the interaction matrix element $V_{\alpha\beta}(\mathbf{k}_1,\mathbf{k}_2,\mathbf{k}_3,\mathbf{k}_4)$ 
that is convenient for numerical calculations is given in Ref.\ \cite{Yoshino19}. 

%************************************************
\subsection{Bogoliubov approximation}\label{sec:BogoliubovTheory_spec}
%************************************************

% [ Bogoliubov Hamiltonian ]--------------------------------------
We now apply the Bogoliubov approximation \cite{Sinova02, Matveenko11, Kwasigroch12, Pethick_Smith_2008}, 
assuming that the condensation occurs at the wave vector $\qv_\alpha$ in the spin-$\alpha$ component. 
To this end, it is useful to introduce
\begin{equation}\label{eq:bt_Vt}
  \bt_{\kv\alpha}:=b_{\qv_\alpha+\kv,\alpha},~~ 
  \Vt_{\alpha\beta} (\kv_1,\kv_2,\kv_3,\kv_4):=V_{\alpha\beta}(\qv_\alpha+\kv_1,\qv_\beta+\kv_2,\qv_\beta+\kv_3,\qv_\alpha+\kv_4).
\end{equation}
By substituting 
\begin{equation}\label{eq:b_cond}
  \bt_{\zerov \alpha} \simeq \bt_{\zerov \alpha}^\dagger \simeq \sqrt{ N_\alpha-\sum_{\kv\ne\zerov} \bt_{\kv\alpha}^\dagger \bt_{\kv\alpha} }
\end{equation}
in Eq.\ \eqref{2ndQ hamiltonian} and retaining terms up to the second order 
in $\bt_{\kv\alpha}$ and $\bt_{\kv\alpha}^\dagger$~($\kv\ne 0$), we obtain the Bogoliubov Hamiltonian
\begin{equation}\label{eq:H_bb}
\begin{split}
  H_\mathrm{int} 
  = &\frac12 \sum_{\alpha,\beta} N_\alpha N_\beta\Vt_{\alpha\beta}(\zerov,\zerov,\zerov,\zerov) 
  - \frac12 \sum_{\kv\ne\zerov}  J (\kv) \\
  &+\frac12 \sum_{\kv\ne\zerov} \left( \bt_{\kv\ua}^\dagger, \bt_{\kv\da}^\dagger, \bt_{-\kv,\ua}, \bt_{-\kv,\da} \right)
  \mathcal{M}(\kv)
 \begin{pmatrix}  \bt_{\kv\ua}\\ \bt_{\kv\da}\\ \bt_{-\kv,\ua}^\dagger\\ \bt_{-\kv,\da}^\dagger \end{pmatrix},
\end{split}
\end{equation}
where
\begin{equation}
\begin{split}
  J (\kv) := \sum_{\alpha,\beta} N_\beta \left[ \Vt_{\alpha\beta}(\kv,\zerov,\zerov,\kv) - \Vt_{\alpha\beta}(\zerov,\zerov,\zerov,\zerov) \right]
  + \sum_\alpha N_\alpha \Vt_{\alpha\alpha} (\kv,\zerov,\kv,\zerov),
  %\lambda_{\alpha\beta} (\kv) := \sqrt{N_\alpha N_\beta} \Vt_{\alpha\beta} (\kv,-\kv,\zerov,\zerov).
\end{split}
\end{equation}
and the expression of the $4\times4$ matrix $\mathcal{M} (\kv)$ is shown in Ref.\ \cite{Yoshino19}. 

% [ Bogoliubov transformation ]--------------------------------------
To diagonalize Eq.\ \eqref{eq:H_bb}, we perform the Bogoliubov transformation
\begin{equation}\label{eq:b_gamma}
  \begin{pmatrix}  \bt_{\kv\ua}\\ \bt_{\kv\da}\\ \bt_{-\kv,\ua}^\dagger\\ \bt_{-\kv,\da}^\dagger \end{pmatrix}
  = W(\kv) 
  \begin{pmatrix}  \gamma_{\kv,1}\\ \gamma_{\kv,2}\\ \gamma_{-\kv,1}^\dagger\\ \gamma_{-\kv,2}^\dagger \end{pmatrix}, ~
W(\kv)=
 \begin{pmatrix}
 \Ucal(\kv) & \Vcal^*(-\kv) \\ \Vcal(\kv) & \Ucal^*(-\kv)
 \end{pmatrix}. 
\end{equation}
Here, the paraunitary matrix $W(\kv)$ is chosen to satisfy 
\begin{equation}\label{eq:tau3MW}
  \tau_3 \mathcal{M}(\kv) W(\kv) = W(\kv) \mathrm{diag} (E_1(\kv),E_2(\kv),-E_1(-\kv),-E_2(-\kv)), 
\end{equation}
where $\tau_3 := \mathrm{diag}(1,1,-1,-1)$. 
Namely, $W(\kv)$ and $E_j(\kv)~(j=1,2)$ are obtained by solving the right eigenvalue problem of $\tau_3\mathcal{M}(\kv)$.
Equation \eqref{eq:H_bb} is then diagonalized as
\begin{equation}\label{eq:H_gg}
 \begin{split}
  H_\mathrm{int} 
  = \frac12 \sum_{\alpha,\beta} N_\alpha N_\beta\Vt_{\alpha\beta}(\zerov,\zerov,\zerov,\zerov) 
  - \frac12 \sum_{\kv\ne\zerov}  J(\kv) 
  + \sum_{\kv\ne\zerov} \sum_{j=1,2} E_j(\kv) \left( \gamma_{\kv j}^\dagger \gamma_{\kv j}+\frac12 \right) .
 \end{split}
\end{equation}
We thus find that the Bogoliubov excitations (bogolons) are created by $\gamma_{\kv j}^\dagger~(j=1,2)$
and have the dispersion relations $\{E_j(\kv) \}$. 

%If the matrix $W(\kv)$ is chosen to satisfy 
%\begin{equation}\label{eq:WMW}
%  W^\dagger (\kv) \mathcal{M}(\kv) W(\kv) = \mathrm{diag} (E_1(\kv),E_2(\kv),E_1(-\kv),E_2(-\kv)),
%\end{equation}
%Therefore, the excitation energies $E_j(\kv)~(j=1,2)$ can be obtained as the right eigenvalues of $\tau_3\mathcal{M}(\kv)$

% [ Ground-state energy ]--------------------------------------
The ground state of Eq.\ \eqref{eq:H_gg} is given by the bogolon vacuum $\ket{0}$, 
which is specified by the condition that $\gamma_{\kv j} \ket{0}=0$ for all $\kv \neq \zerov$ and $j=1,2$. 
The ground-state energy $E_\GS$ (scaled by the interaction energy scale $gn^2A$) is therefore given by 
\begin{equation}\label{eq:Egs/atom}
\begin{split}
 \frac{E_\GS}{gn^2 A} 
 =& \frac12 \sum_{\alpha,\beta} \frac{A}{g} \Vt_{\alpha\beta}(\zerov,\zerov,\zerov,\zerov) 
 +\frac{1}{gn \nu \Nvor}\sum_{\kv\ne\zerov}
 \left[\sum_j E_j(\kv) - J(\kv) \right]  \\
 =& \frac12 \sum_{\alpha,\beta} \frac{A}{g} \Vt_{\alpha\beta}(\zerov,\zerov,\zerov,\zerov) 
 + \frac{1}{2\pi gn \nu} \int_\mathrm{BZ} d^2\kv \ell^2 
 \left[ \sum_j E_j(\kv) - J(\kv)  \right] .
\end{split}
\end{equation}
Here, in the final expression, we take the thermodynamic limit $\Nvor\to\infty$ so that the sum is replaced by the integral over the Brillouin zone as
$\frac{1}{\Nvor} \sum_{\kv\ne\zerov}\to \frac{1}{|\bv_1\times\bv_2|} \int \drm^2\kv = \frac1{2\pi} \int_\mathrm{BZ} \drm^2\kv\ell^2$; 
this integral is convergent as the integrand is finite over the entire Brillouin zone. 
The first term of Eq.\ \eqref{eq:Egs/atom} corresponds to the mean-field ground-state energy, 
which has been analyzed by Mueller and Ho \cite{Mueller02}. 
The other term gives a quantum correction and is inversely proportional to the filling factor $\nu$. 
In Sec.\ \ref{sec:phase_diagrams}, we numerically calculate Eq.\ \eqref{eq:Egs/atom}, 
and discuss how the quantum correction affects the ground-state phase diagrams. 

% [ correlators and depletion ]--------------------------------------
Using Eq.\ \eqref{eq:b_gamma}, we can calculate the following correlators in the ground state: 
%In order to discuss the intercomponent entanglement, we calculate the nonzero two-point correlation functions of the ground state given by 
\begin{subequations}\label{eq:2 point corr. 2comp}
\begin{align}
 &\bra{0}\bt_{\kv,\alpha}^\dagger\bt_{\kv,\alpha}\ket{0} 
 = \sum_j |\Vcal_{\alpha,j}(-\kv)|^2,~~
 \bra{0}\bt_{-\kv,\alpha}\bt_{-\kv,\alpha}^\dagger\ket{0} 
 = \sum_j |\Ucal_{\alpha,j}(-\kv)|^2, \label{eq:depletion} \\
 &\bra{0}\bt_{-\kv,\alpha}\bt_{\kv,\alpha}\ket{0} 
 = \sum_j \Ucal_{\alpha,j}(-\kv) \Vcal_{\alpha,j}^\ast(-\kv),~~
 \bra{0}\bt_{\kv,\alpha}^\dagger\bt_{-\kv,\alpha}^\dagger\ket{0} 
 = \sum_j \Ucal_{\alpha,j}^\ast(-\kv) \Vcal_{\alpha,j}(-\kv). \label{eq:anomalous}
\end{align}
\end{subequations}
Here, we have nonzero ``anomalous'' correlators in Eq.\ \eqref{eq:anomalous} as the particle numbers $N_\ua$ and $N_\da$ are not conserved in the Bogoliubov Hamiltonian \eqref{eq:H_bb}. 
%where the ``anomalous" correlation functions \eqref{eq:anomalous} are nonzero in the present case in contrast to (something).
Using Eq.\ \eqref{eq:depletion}, we can further calculate the fraction of depletion $n'/n$, which is equal for the two components, as
\begin{equation}\label{eq:depletion_Vcal}
 \frac{n'}{n}
 =\frac{1}{N_\alpha}\sum_{\kv\ne\zerov} \bra{0}\bt_{\kv,\alpha}^\dagger\bt_{\kv,\alpha}\ket{0} 
 = \frac{2}{\nu\Nvor} \sum_{\kv\ne\zerov} \sum_{j=1,2} |\Vcal_{\alpha,j}(-\kv)|^2~~(\alpha=\ua,\da).
\end{equation}
As discussed in Sec.\ \ref{sec:EFT_vorlat_corr} [see Eq.\ \eqref{eq:dep_log}], this quantity is expected to diverge logarithmically as a function of $\Nvor$. 
We will confirm this behavior numerically in Sec.\ \ref{sec:numerics_dep}. 
This diverging behavior comes from the divergence of $|\Vcal_{\alpha,2}(-\kv)|^2$ for $\kv\to\zerov$ in Eq.\ \eqref{eq:depletion_Vcal}. 
%In contrast, in the ground-state energy density in Eq.\ \eqref{eq:Egs/atom}, all the quantities in the summand are finite over the entire Brillouin zone, 
%leading to a convergent result even in the limit $\Nvor\to\infty$. 
We note that the Bogoliubov theory should be applied in the condition of weak depletion $n'/n\ll 1$; 
this condition is satisfied in typical experiments of ultracold atomic gases, 
where $\Nvor$ is at most of the order of 100 \cite{Schweikhard04_1comp}. 
%Despite the diverging behavior of $n'/n$, the ground state energy density in Eq.\ \eqref{eq:Egs/atom} converges for sufficiently large $\Nvor$ 
%as all the quantities in the summand are finite over the entire Brillouin zone. 

%which is obtained by the Bogoliubov translation \eqref{eq:BogoTrnf} diverges as $n_{\mathrm{dep}} \sim \ln(\Nvor)/\nu$ 
%$\frac1N \sum_{\kv\ne\zerov,\alpha} \langle\bt_{\kv\alpha}^\dagger\bt_{\kv\alpha}\rangle \sim \ln(\Nvor)/\nu$ 
%\cite{Sinova02,Kwasigroch12} in the thermodynamic limit where the Bogoliubov approximation is not valid. 
%The Bogoliubov theory is still applicable since $\Nvor$ is at most of the order of 100 in typical experiments of ultracold atomic gases \cite{Schweikhard04_1comp}. 
% ### Calculate the depletions with different values of the filling factor. 

%The ground state energy is given by $E_\GS=\bra{\GS}H_\mathrm{int}\ket{\GS}$ where the ground state $\ket{\GS}$ is specified by the condition $\gamma_{\kv i} \ket{\GS}=0$ for all $\kv \neq \zerov$ and $i=1,2$. 
%The dimensionless ground state energy with the filling factor $\nu$, which we numerically calculate, is given by 
%Specifying the ground state $\ket{\GS}$ by the condition $\gamma_{\kv i} \ket{\GS}=0$ for all $\kv \neq \zerov$ and $i=1,2$, the ground-state energy $E_\GS=\bra{\GS}H_\mathrm{int}\ket{\GS}$ is given as
%In the thermodynamics limit $\Nvor\to\infty$, we replace the summation in Eq.~\eqref{eq:Egs/atom} by $1/\Nvor \sum_{\kv\ne\zerov}\to 1/|\bv_1\times\bv_2| \int d^2\kv = \frac1{2\pi} \int d^2\kv\ell^2$.

%************************************************
\subsection{Intercomponent entanglement}\label{sec:BogoliubovTheory_ent}
%************************************************

% [ RDM ]--------------------------------------
For the RDM $\rho_\ua$ for the spin-$\ua$ component, 
we introduce the following Gaussian ansatz \cite{Chen13,Lundgren13,Peschel03,Peschel09}: 
\begin{equation}\label{eq:He_Mbb}
 \rho_\ua=\frac1{Z_\erm} \erm^{-H_\erm},~~
 H_\erm= \frac12 \sum_{\kv\ne\zerov} \left(\bt_{\kv,\ua}^\dagger,\bt_{-\kv,\ua}\right) 
 M_\erm(\kv)\begin{pmatrix} \bt_{\kv,\ua} \\ \bt_{-\kv,\ua}^\dagger \end{pmatrix} ,~~
 Z_\erm= \Tr~\erm^{-H_\erm}, 
\end{equation} 
with
\begin{equation}
 M_\erm (\kv)
 =\begin{pmatrix} h_\kv& -\lambda_\kv\\ -\lambda_{-\kv}^\ast& h_{-\kv} \end{pmatrix},~
 \lambda_\kv=\lambda_{-\kv}. 
\end{equation}
By performing a Bogoliubov transformation 
\begin{equation}\label{eq:Bogo. trnf. of RDM}
 \begin{pmatrix} \bt_{\kv\ua} \\ \bt_{-\kv,\ua}^\dagger \end{pmatrix} = W_\erm(\kv) 
 \begin{pmatrix} \eta_\kv \\ \eta_{-\kv}^\dagger \end{pmatrix} ,~~
 W_\erm (\kv) 
 =\begin{pmatrix} 
  \cosh\theta_\kv & \erm^{-\irm\phi_\kv}\sinh\theta_\kv \\ 
  \erm^{\irm\phi_\kv}\sinh\theta_\kv & \cosh\theta_\kv \end{pmatrix} 
\end{equation}
with 
\begin{equation}
 \cosh2\theta_\kv=\frac{h_\kv+h_{-\kv}}{2\tilde{h}_\kv},~~
 \erm^{-i\phi_\kv} \sinh2\theta_\kv=\frac{\lambda_\kv}{\tilde{h}_\kv}, ~~
 \tilde{h}_\kv=\sqrt{\frac14\left(h_\kv+h_{-\kv}\right)^2 - |\lambda_\kv|^2} ,
\end{equation}
the entanglement Hamiltonian in $H_\erm$ in Eq.\ \eqref{eq:He_Mbb} is diagonalized as
\begin{equation}
 H_\erm=\frac12 \sum_{\kv\ne\zerov}\left(\xi_\kv\eta_\kv^\dagger\eta_\kv+\xi_{-\kv}\eta_{-\kv}\eta_{-\kv}^\dagger \right)
 =\sum_{\kv\ne\zerov}\xi_\kv\left(\eta_\kv^\dagger \eta_\kv+\frac12\right) ,
\end{equation}
where
\begin{equation}
 \xi_\kv:=\tilde{h}_\kv+\frac{h_\kv-h_{-\kv}}{2} 
\end{equation}
is the single-particle ES. 

% [ Correlators and Bose distribution function ]--------------------------------------
Using the relation \eqref{eq:Bogo. trnf. of RDM} and the Bose distribution functions
\begin{equation}
 %\langle \eta_\kv^\dagger \eta_\kv \rangle =
 \Tr\left( \eta_\kv^\dagger \eta_\kv \rho_\ua \right) 
 =\frac{1}{\erm^{\xi_\kv}-1} = f_\Brm(\xi_\kv) ,~~
  \Tr\left( \eta_{-\kv} \eta_{-\kv}^\dagger \rho_\ua \right) = 1+f_\Brm(\xi_{-\kv})=-f_\Brm(-\xi_{-\kv}),
\end{equation}
we obtain 
\begin{subequations}\label{eq:2 point corr. 1comp}
\begin{align}
 \Tr\left( \bt_{\kv,\ua}^\dagger\bt_{\kv,\ua} \rho_\ua \right) 
 &=f_\Brm (\xi_\kv)\cosh^2\theta_\kv - f_\Brm(-\xi_{-\kv}) \sinh^2\theta_\kv ,  \label{eq:b^dag b up} \\
 \Tr\left( \bt_{-\kv,\ua}\bt_{-\kv,\ua}^\dagger \rho_\ua \right) 
 &=-f_\Brm(-\xi_{-\kv}) \cosh^2\theta_\kv +f_\Brm(\xi_\kv)\sinh^2\theta_\kv ,  \label{eq:b b^dag up} \\
 2\Tr\left( \bt_{-\kv,\ua}\bt_{\kv,\ua} \rho_\ua \right) 
 &%=2\left[f_\Brm(\xi_\kv)-f_\Brm(-\xi_{-\kv})\right]\erm^{-i\phi_\kv}\cosh\theta_\kv\sinh\theta_\kv
   =  \left[f_\Brm(\xi_\kv)-f_\Brm(-\xi_{-\kv})\right]\erm^{-i\phi_\kv}\sinh\left(2\theta_\kv\right) .  \label{eq:bb up}
\end{align}
\end{subequations}
We require these to be equal to the correlators \eqref{eq:2 point corr. 2comp} with respect to the Bogoliubov ground state. 
We can then express $f_\Brm(\xi_\kv)$ in terms of the correlators \eqref{eq:2 point corr. 2comp} in the following way. 
First, by taking the sum and the difference of Eqs.~\eqref{eq:b^dag b up} and \eqref{eq:b b^dag up}, we have
\begin{subequations}
\begin{align}
 \langle\bt_{\kv\ua}^\dagger\bt_{\kv\ua}\rangle + \langle\bt_{-\kv,\ua}\bt_{-\kv,\ua}^\dagger\rangle
 &=\left[ f_\Brm(\xi_\kv)-f_\Brm(-\xi_{-\kv}) \right] \cosh \left(2\theta_\kv \right) , 
 \label{eq:b^dag b+b b^dag} \\
 \langle \bt_{\kv\ua}^\dagger\bt_{\kv\ua}\rangle - \langle\bt_{-\kv,\ua}\bt_{-\kv,\ua}^\dagger\rangle
 &= f_\Brm(\xi_\kv) +f_\Brm(-\xi_{-\kv})  \label{eq:b^dag b-b b^dag} ,
\end{align}
\end{subequations}
where we take the shorthand notation $\langle\cdot\rangle:=\bra{0}\cdot\ket{0}$.  
Next, using Eqs.\ \eqref{eq:bb up} and \eqref{eq:b^dag b+b b^dag}, we have
\begin{equation}\label{eq:f(k)+f(^k)+1} 
 f_\Brm (\xi_\kv)-f_\Brm (-\xi_{-\kv})
 =\sqrt{\left(\langle\bt_{\kv\ua}^\dagger\bt_{\kv\ua}\rangle
 +\langle\bt_{-\kv,\ua}\bt_{-\kv,\ua}^\dagger\rangle\right)^2
 - 4\big|\langle\bt_{-\kv,\ua}\bt_{\kv\ua}\rangle\big|^2} . 
\end{equation}
Lastly, using Eqs.~\eqref{eq:b^dag b-b b^dag} and \eqref{eq:f(k)+f(^k)+1}, we obtain 
\begin{equation}\label{eq:BoseDistributionFunction}
 f_\Brm(\xi_\kv) 
 =\frac12\left(\corr{\bt_{\kv\ua}^\dagger\bt_{\kv\ua}}- \corr{\bt_{-\kv,\ua}\bt_{-\kv,\ua}^\dagger}\right)
 +\sqrt{\frac14\left(\corr{\bt_{\kv\ua}^\dagger\bt_{\kv\ua}}+\corr{\bt_{-\kv,\ua}\bt_{-\kv,\ua}^\dagger} \right)^2
 - \big|\corr{\bt_{-\kv,\ua}\bt_{\kv\ua} }\big|^2} ,
\end{equation}
from which we can calculate the single-particle ES $\xi_\kv = \ln \left[ 1+f_B(\xi_\kv)^{-1} \right]$.

% [ ES and EE ]--------------------------------------
We can further use Eq.\ \eqref{eq:BoseDistributionFunction} to calculate the intercomponent EE as
\begin{equation}\label{eq:EE}
 S_\erm=\sum_{\kv\ne\zerov} \left\{-f_\Brm(\xi_\kv) \ln f_\Brm(\xi_\kv)+\left[1+f_\Brm(\xi_\kv)\right]\ln\left[1+f_\Brm(\xi_\kv)\right]\right\}.
\end{equation}
As discussed in Sec.\ \ref{sec:ent_vorlat} [see Eq.\ \eqref{eq:Se_Nv_logNv}], $S_\erm$ is expected to show a volume-law behavior with a subleading logarithimic correction. 
The EE per flux quantum in the thermodynamic limit is then expressed in the integral form
\begin{equation}\label{eq:EE_Nvor}
  \lim_{\Nvor\to\infty} \frac{S_\erm}{\Nvor}=\frac{1}{2\pi} \int_\mathrm{BZ} \drm^2\kv\ell^2 \left\{-f_\Brm(\xi_\kv) \ln f_\Brm(\xi_\kv)+\left[1+f_\Brm(\xi_\kv)\right]\ln\left[1+f_\Brm(\xi_\kv)\right]\right\}.
\end{equation}
We note that the correlators \eqref{eq:2 point corr. 2comp} are independent of $\nu$ once the lattice structure is fixed. 
Therefore, the EE per flux quantum in Eq.\ \eqref{eq:EE_Nvor} is also independent of $\nu$ in a similar manner. 
In Sec.\ \ref{sec:numerics_ent}, we will calculate the EE per flux quantum assuming the structure in the mean-field ground state. 
%(if the lattice structure changes due to the quantum correction, the EE per flux quantum also changes accordingly). 

%%%%%%%%%%%%%%%%%%%%%%%%%%%%%%%%%%%%%%%%%%%%%%%%%
\section{Numerical results}\label{sec:NumericalResults}
%%%%%%%%%%%%%%%%%%%%%%%%%%%%%%%%%%%%%%%%%%%%%%%%%

In this section, we present numerical results that are obtained using the Bogoliubov theory formulation of Sec.\ \ref{sec:BogoliubovTheory}. 
In this formulation, one starts from the lattice structure 
with the primitive vectors $\av_1$ and $\av_2$ and the displacement parameters $u_1$ and $u_2$ [see Fig.\ \ref{fig:Phases_GP}(f)]. 
In the GP mean-field theory, these parameters are determined so as 
to minimize the mean-field ground-state energy [the first term on the right-hand side of Eq.\ \eqref{eq:Egs/atom}] for a fixed magnetic length $\ell$. 
As demonstrated by Mueller and Ho \cite{Mueller02}, this mean-field analysis gives a rich phase diagram 
that consists of five different %lattice structures 
vortex-lattice phases 
as shown in Fig.\ \ref{fig:Phases_GP}. 
In Secs.\ \ref{sec:numerics_renorm}, \ref{sec:numerics_excitation}, \ref{sec:numerics_ent}, and \ref{sec:numerics_dep}, 
we analyze renormalized coupling constants, excitation spectra, intercomponent entanglement, and the fraction of depletion, respectively, 
using the Bogoliubov theory based on the mean-field vortex lattice structures. 
Therefore, the results in these sections correspond to the case of $\nu=\infty$. 
As we lower the filling factor $\nu$, quantum fluctuations are expected to affect the vortex lattice structures and the ground-state phase diagrams. 
In Sec.\ \ref{sec:phase_diagrams}, we investigate how quantum fluctuations affect the ground-state phase diagrams for parallel and antiparallel fields 
by calculating a quantum correction to the ground-state energy [i.e., the term proportional to $\nu^{-1}$ in Eq.\ \eqref{eq:Egs/atom}]. 

%************************************************
\subsection{Renormalized coupling constants}\label{sec:numerics_renorm}
%************************************************

%############################
\begin{figure}
 \centering
     \includegraphics[width=8cm, angle=0]{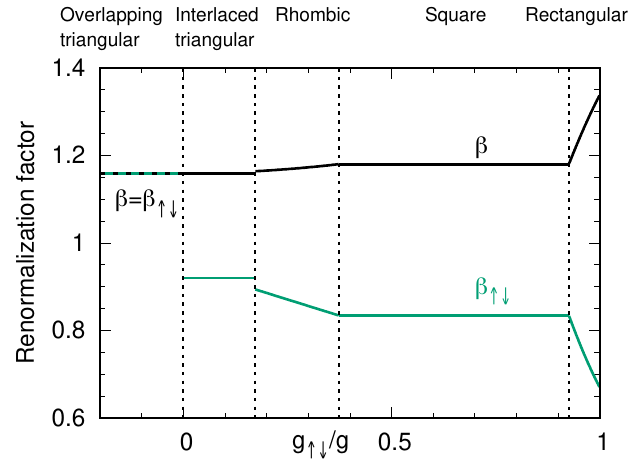}
 \caption{\label{fig:renorm_int}
Renormalization factors $\beta:=\gb/g$ (black) and $\beta_{\ua\da}:=\gb_{\ua\da}/g_{\ua\da}$ (green) for the intracomponent and intercomponent coupling constants.  
These factors are calculated using Eq.\ \eqref{eq:g_renorm_LLL} for the mean-field vortex lattice structure for each $g_{\ua\da}/g$, 
and take the same values for parallel and antiparallel fields. 
We set $\Nvor=97^2$, with which a sufficient convergence to the thermodynamic limit is achieved. 
For overlapping triangular lattices ($-1<g_{\ua\da}/g<0$), we have $\beta=\beta_{\ua\da}$ and thus the two curves overlap; 
the value $\beta=\beta_{\ua\da}=1.1596$ in this region coincides with the renormalization factor obtained for a scalar BEC \cite{Fetter09,Sinova02,Aftalion05}. 
Vertical dashed lines indicate the transition points in the mean-field phase diagram \cite{Mueller02}. 
}
\end{figure}
%############################

% [ Expression  ]--------------------------------------
We first determine the renormalized coupling constants $\gb_{\alpha\beta}$ 
or equivalently, the renormalization factors $\beta_{\alpha\beta}:=\gb_{\alpha\beta}/g_{\alpha\beta}$ 
that are introduced in Sec.\ \ref{sec:EFT_vorlat_derive}. 
Comparing the mean-field ground-state energy [the first term of Eq.\ \eqref{eq:Egs/atom}] 
with the corresponding field-theoretical expression \eqref{eq:EGS_fieldtheory}, we find
\begin{equation}\label{eq:g_renorm_LLL}
 \gb_{\alpha\beta} = \beta_{\alpha\beta} g_{\alpha\beta} =  A V_{\alpha\beta}(\qv_\alpha,\qv_\beta,\qv_\beta,\qv_\alpha) 
   =Ag_{\alpha\beta} \int \drm^2\mathbf{r}~
  |\Psi_{\qv_\alpha, \alpha}(\mathbf{r})|^2  |\Psi_{\qv_\beta, \beta}(\mathbf{r})|^2.
\end{equation}
This expression can also be obtained by substituting the condensate wave function $\sqrt{N_\alpha}\Psi_{\qv_\alpha, \alpha}(\mathbf{r})$ into $\psi_\alpha(\rv)$ in Eq.\ \eqref{eq:g_renorm}. 
As seen in this expression, $\gb_{\alpha\beta}$ is determined from the contribution of each interaction term to the mean-field ground-state energy. 

% [ Result  ]--------------------------------------
Figure \ref{fig:renorm_int} shows the renormalization factors $\beta:=\beta_{\ua\ua}=\beta_{\da\da}$ and $\beta_{\ua\da}=\beta_{\da\ua}$ 
calculated for the mean-field vortex lattice structures. 
For overlapping triangular, interlaced triangular, and square lattices, 
$\beta$ and $\beta_{\ua\da}$ do not depend on $g_{\ua\da}/g$ as the lattice structures remain unchanged in the concerned regions. 
For rhombic and rectangular lattices, in contrast, $\beta$ and $\beta_{\ua\da}$ do depend on $g_{\ua\da}/g$ 
as the inner angle $\theta$ and the aspect ratio $b/a$ continuously vary for the former and the latter, respectively. 

We have argued in Sec.\ \ref{sec:EFT_vorlat_derive} that  the intracomponent coupling is always enhanced by the renormalization.  
We indeed find $\beta>1$ in all the regions in Fig.\ \ref{fig:renorm_int}. 
We have also argued that the intercomponent repulsion $g_{\ua\da}>0$ (attraction $g_{\ua\da}<0$) is reduced (enhanced) by the renormalization 
owing to the displacement (overlap) of vortices between the components. 
In Fig.\ \ref{fig:renorm_int}, we indeed find $\beta_{\ua\da}<1$ ($\beta_{\ua\da}>1$) for $g_{\ua\da}>0$ ($g_{\ua\da}<0$). 
Furthermore, $\beta$ ($\beta_{\ua\da}$) monotonically increases (decreases) as a function of $g_{\ua\da}/g$ for $g_{\ua\da}>0$.  
This reflects the fact that with increasing $g_{\ua\da}/g$, vortices in different components tend to repel more strongly with each other 
at the cost of increasing the intracomponent interaction energy.

\subsection{Excitation spectrum and elastic constants}\label{sec:numerics_excitation}
%************************************************

%############################
\begin{figure*}
 \centering
    \includegraphics[width=14cm, angle=0]{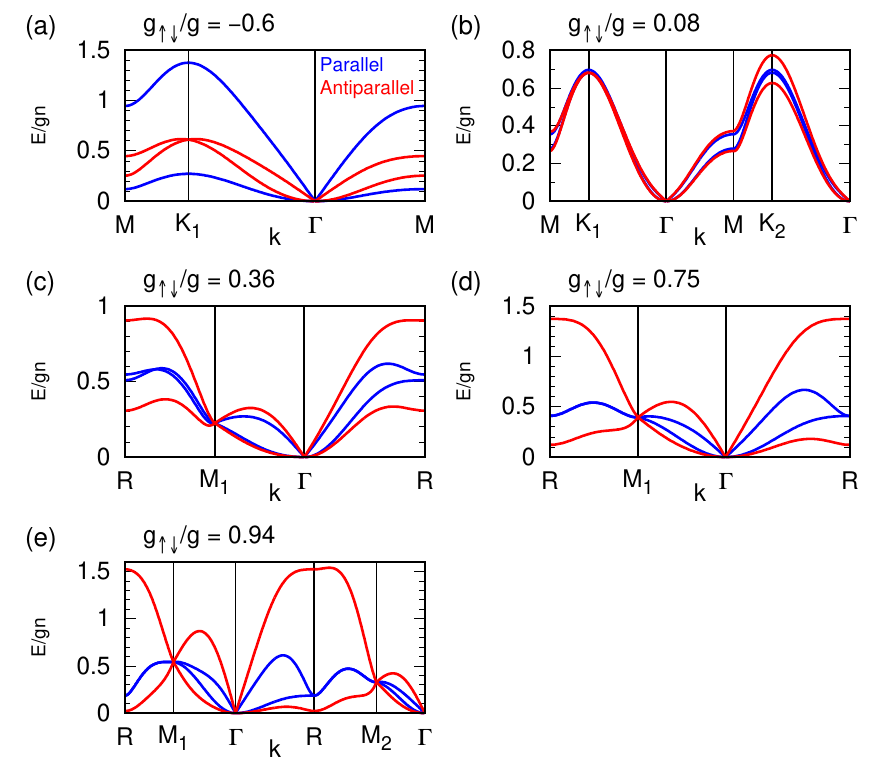}%{hsp_unrescale.pdf}
 \caption{\label{fig:spec_unresc}
 Bogoliubov excitation spectra $\{E_j(\kv)\}$ (in units of $gn$) for different values of $g_{\ua\da}/g$ 
 corresponding to the lattice structures (a)-(e) shown in Fig.\ \ref{fig:Phases_GP}. 
 % (a) overlapping triangular, (b) interlaced triangular, (c) rhombic, (d) square, and (e) rectangular lattices. 
 %Each panel shows both results for parallel (blue) and antiparallel (red) magnetic fields. 
 In each panel, both results for parallel (blue) and antiparallel (red) magnetic fields are shown.
 Calculations are done along %the paths %indicated by 
 dotted arrows in the lower panels of Fig.\ \ref{fig:Phases_GP}. 
 }
\end{figure*}
%############################

%############################
\begin{figure*}
 \centering
    \includegraphics[width=14cm, angle=0]{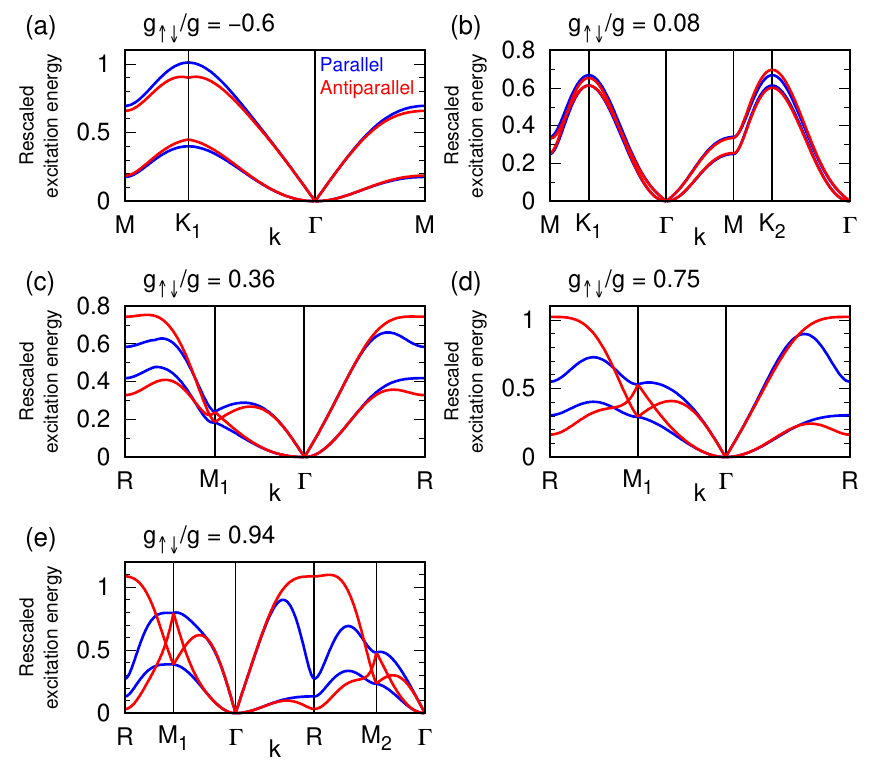}%{hsp_v2.pdf}
 \caption{\label{fig:spec_resc}
 Rescaled Bogoliubov excitation spectra for the five cases (a)-(e) shown in Fig.\ \ref{fig:spec_unresc}.
 %for different values of $g_{\ua\da}/g$ corresponding to the lattice structures (a)-(e) shown in Fig.\ \ref{fig:Phases_GP}: 
 %(a) overlapping triangular, (b) interlaced triangular, (c) rhombic, (d) square, (e) rectangular lattices. 
 Blue curves show $E_1^\mathrm{P}(\kv)/(\sqrt{g\gb_-}n)$ and $E_2^\mathrm{P}(\kv)/(\sqrt{g\gb_+}n)$ for parallel (P) fields 
 while red curves show $E_1^\mathrm{AP}(\kv)/(\sqrt{g\gb_+}n)$ and $E_2^\mathrm{AP}(\kv)/(\sqrt{g\gb_-}n)$ for antiparallel (AP) fields. 
 Here, $E_j^\mathrm{P/AP}(\kv)$ with $j=1$ and $2$ correspond to the upper and lower excitation bands, respectively. 
 We can confirm that the blue and red curves overlap at sufficiently low energies around the $\Gamma$ point,  
 which indicates the rescaling relations in Eq.\ \eqref{eq:E12_rescaling}. 
 }
\end{figure*}
%############################

%############################
\begin{figure*}
 \centering
    \includegraphics[width=12cm, angle=0]{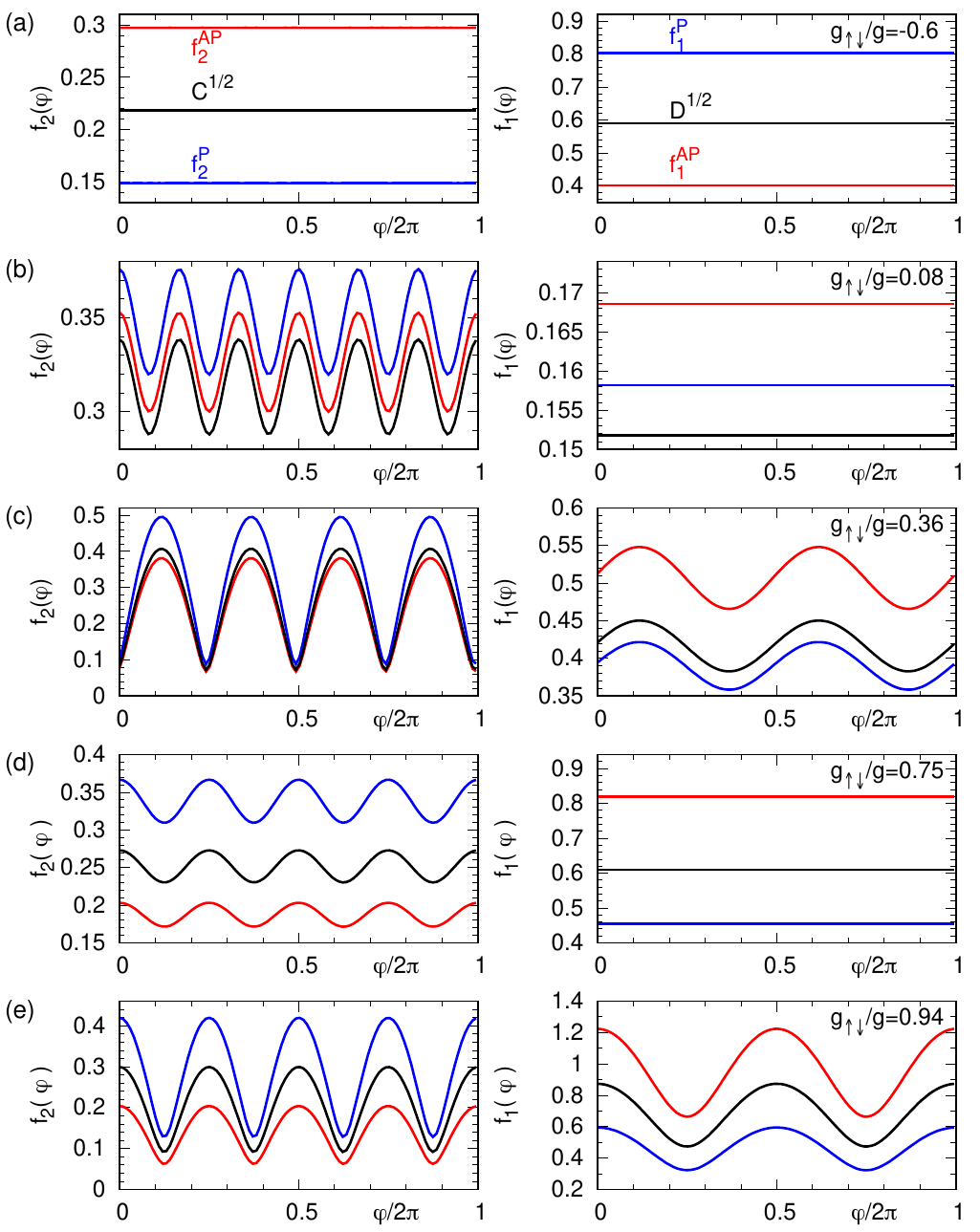}%{aniso.pdf}
 \caption{\label{fig:aniso_energy}
Dimensionless functions $f_2^{\mathrm{P}/\mathrm{AP}}(\varphi)$ (left) and $f_1^{\mathrm{P}/\mathrm{AP}}(\varphi)$ (right) 
for parallel ($\mathrm{P}$; blue) and antiparallel ($\mathrm{AP}$; red) fields 
for the same cases (a)-(e) as in Figs.\ \ref{fig:spec_unresc} and \ref{fig:spec_resc}. 
These functions express the anisotropy of the low-energy spectra as in Eq.\ \eqref{eq:e_k_fphi}, 
and they are calculated from the %Bogoliubov excitation spectra 
$\{E_j(\kv)\}$ along a circular path $\kv=k(\cos\varphi,\sin\varphi)$ with $k=0.001a/\ell^2$ and $\varphi \in [0,2\pi)$. 
With proper rescaling as in Eq.\ \eqref{eq:energy_rescaling}, the curves for parallel and antiparallel fields are found to agree perfectly up to numerical precision. 
Specifically, in the left panels, $f_2^\mathrm{P}(\varphi)\sqrt{g/\gb_+}$ and $f_2^\mathrm{AP}(\varphi)\sqrt{g/\gb_-}$ give the common function $\sqrt{C(\varphi)}$ (black). 
In the right panels, $f_1^\mathrm{P}(\varphi)\sqrt{g/\gb_-}$ and $f_1^\mathrm{AP}(\varphi)\sqrt{g/\gb_+}$ give the common function $\sqrt{D(\varphi)}$ (black). 
}
\end{figure*}
%############################

\begin{figure*}
\centering\includegraphics[width=12cm, angle=0]{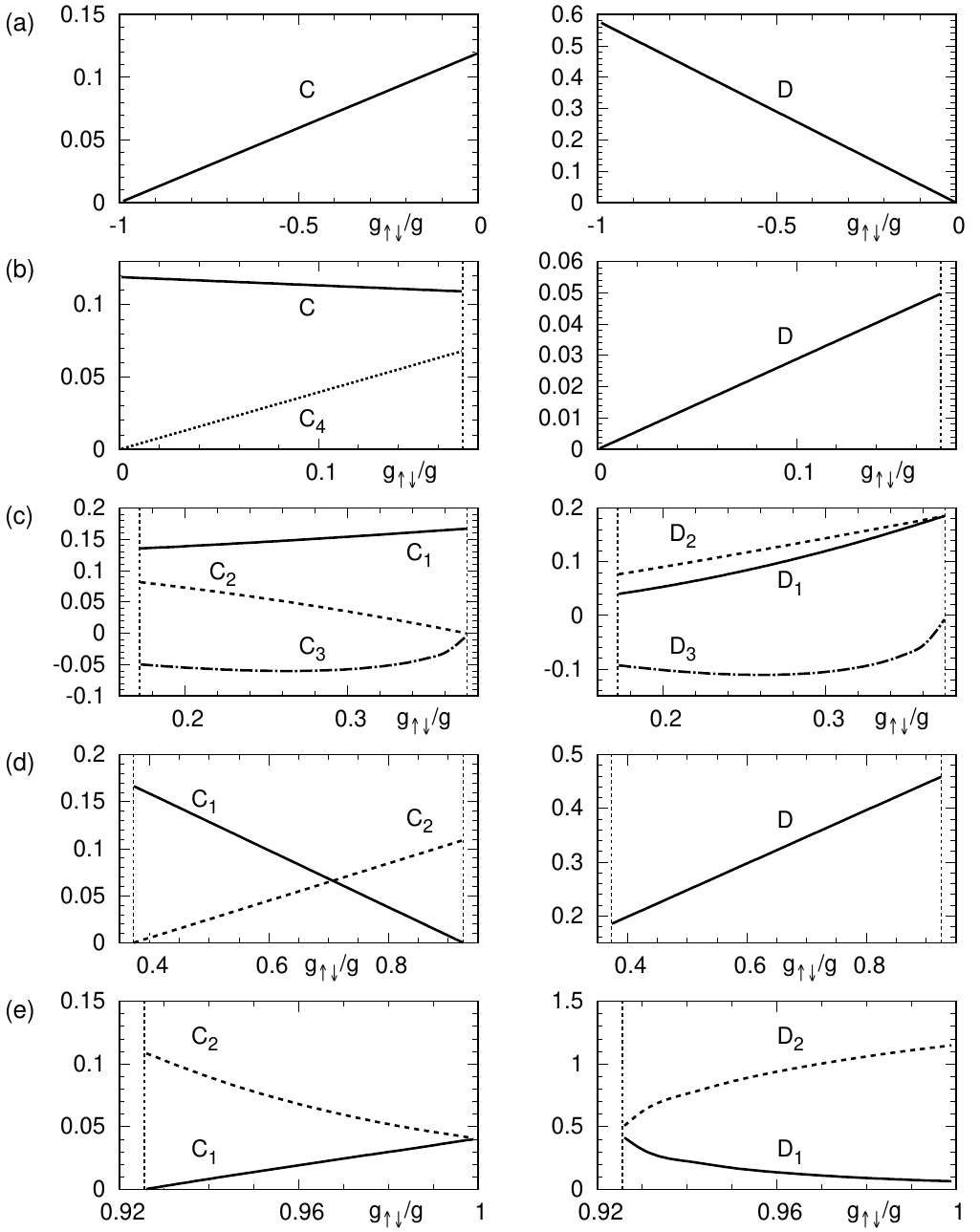}%{elcon_200703.pdf}
\caption{\label{fig:elacon}
 Dimensionless elastic constants $C_i~(i=1,2,3,4)$ (left) and $D_i~(i=1,2,3)$ (right) for the lattice structures (a)-(e) shown in Fig.\ \ref{fig:Phases_GP}.
 %(a) overlapping triangular, (b) interlaced triangular, (c) rhombic, (d) square, and (e) rectangular lattices.  
 These constants are obtained by fitting the numerically obtained functions $\sqrt{C(\varphi)}$ and $\sqrt{D(\varphi)}$ 
(common for parallel and antiparallel fields as shown in Fig.\ \ref{fig:aniso_energy}) using Eq.\ \eqref{eq:CD_phi}. 
 See Eq.\ \eqref{eq:ElasEngy_ElasConst} for the symmetry constraints on the constants. 
 Vertical dashed lines indicate the transition points in the mean-field ground state.
 }
\end{figure*}

% [ How to obtain spectrum; comparison with previous work ]--------------------------------------
As explained in Sec.\ \ref{sec:BogoliubovTheory_spec} [see Eq.\ \eqref{eq:tau3MW}], the excitation spectrum $E_j(\kv)~(j=1,2)$ can be obtained 
by numerically calculating the right eigenvalues of the $4\times4$ matrix $\tau_3\Mcal(\kv)$. 
Figure \ref{fig:spec_unresc} presents spectra obtained in this way 
for all the lattice structures (a)-(e) shown in Fig.\ \ref{fig:Phases_GP} and for both parallel and antiparallel fields. 
In Ref.\ \cite{Yoshino19},\footnote{
There are errors in the scales of some figures in Ref.\ \cite{Yoshino19}. 
Specifically, the numerical data for the vertical axes in Figs.\ 2, 4, and 5 should be multiplied by $1/4$, $1/\left(2\sqrt{2}\right)$, and $1/8$, respectively. 
These errors are unrelated to the issue of renormalization discussed in the present paper. 
%Fig.\ 2: $4E/(gn)$. Fig.\ 3: $2\sqrt{2}f_j(\varphi)$. Fig.\ 
As our understanding of the rescaling relations is now updated from Ref.\ \cite{Yoshino19}, 
Figs.\ \ref{fig:spec_unresc}, \ref{fig:aniso_energy}, and \ref{fig:elacon} of the present paper could be seen as %substitutes for 
improved versions of 
these figures. 
}  
we discuss various unique features of these spectra 
such as linear and quadratic dispersion relations at low energies 
and the emergence of line and point nodes at high energies that are related to a fractional translation symmetry. 
Here, we aim to demonstrate the rescaling relations in Eqs.\ \eqref{eq:E12_rescaling} and \eqref{eq:energy_rescaling} 
which are predicted by the low-energy effective field theory. 
In Ref.\ \cite{Yoshino19}, the unrenormalized coupling constants $g_{\alpha\beta}$ are used for the rescaling, 
which leads to an incorrect conclusion that the rescaling relations hold only for overlapping triangular lattices. 
Using the renormalized coupling constants obtained in Sec.\ \ref{sec:numerics_renorm}, 
we can demonstrate the rescaling relations for all the five structures (a)-(e) shown in Fig.\ \ref{fig:Phases_GP}. 

% [ Results on spectra ]--------------------------------------
Figure \ref{fig:spec_resc} displays rescaled excitation spectra for the five cases (a)-(e) in Fig.\ \ref{fig:spec_unresc}. 
Here, $\gb_\pm:=\gb\pm \gb_{\ua\da}=\beta g\pm \beta_{\ua\da} g_{\ua\da}$ are used for the rescaling, 
where the renormalization factors $\beta$ and $\beta_{\ua\da}$ are shown in Fig.\ \ref{fig:renorm_int}.
We can confirm that the rescaling relations in Eq.\ \eqref{eq:E12_rescaling} hold at sufficiently low energies around the $\Gamma$ point. 
Interestingly, in Figs.\ \ref{fig:spec_resc}(a) and (b), the rescaling relations hold approximately up to high energies, 
which is beyond the scope of effective field theory. 
At low energies, the spectra in Fig.\ \ref{fig:spec_unresc} can be fit well by linear and quadratic dispersion relations 
\begin{equation}\label{eq:e_k_fphi}
 E_j(\kv) = \sqrt2 gn(k\ell)^j f_j(\varphi)~~(j=1,2) ,
\end{equation}
where the wave vector is parametrized as $\kv=k(\cos\varphi,\sin\varphi)$ and $\{f_j(\varphi)\}$ are dimensionless functions that characterize the anisotropy of the spectrum. 
Figure \ref{fig:aniso_energy} shows the functions $\{f_j(\varphi)\}$ obtained numerically for the same cases as in Figs.\ \ref{fig:spec_unresc} and \ref{fig:spec_resc}. 
We find that with proper rescaling as in Eq.\ \eqref{eq:energy_rescaling}, the curves for parallel and antiparallel fieds coincide perfectly up to numerical precision, 
giving the funcions $\sqrt{C(\varphi)}$ and $\sqrt{D(\varphi)}$ that are related to the elastic constants.  

%############################
%############################

% [ Elastic constants ]--------------------------------------
By comparing the obtained $\sqrt{C(\varphi)}$ and $\sqrt{D(\varphi)}$ with the analytical expressions in Eq.\ \eqref{eq:CD_phi}, 
we can determine the dimensionless elastic constants $\{C_i\}$ and $\{D_i\}$. 
Figure \ref{fig:elacon} shows the determined elastic constants as functions of $g_{\ua\da}/g$, 
which are common for parallel and antiparallel fields. 
In our previous work \cite{Yoshino19}, we obtained different elastic constants for the two types of fields 
as we missed the necessity of using the renormalized coupling constants in relating the spectra to the elastic constants. 
Figure \ref{fig:elacon} is essentially consistent with the elastic constants determined by a different method by Ke\c{c}eli and Oktel \cite{Oktel06} 
except that the constant $C_4$ in interlaced triangular lattices (b) was missed in Ref.\ \cite{Oktel06}. 

%************************************************
\subsection{Intercomponent entanglement spectrum and entropy}\label{sec:numerics_ent}
%************************************************

%############################
\begin{figure*}
\centering\includegraphics[width=12cm , angle=0]{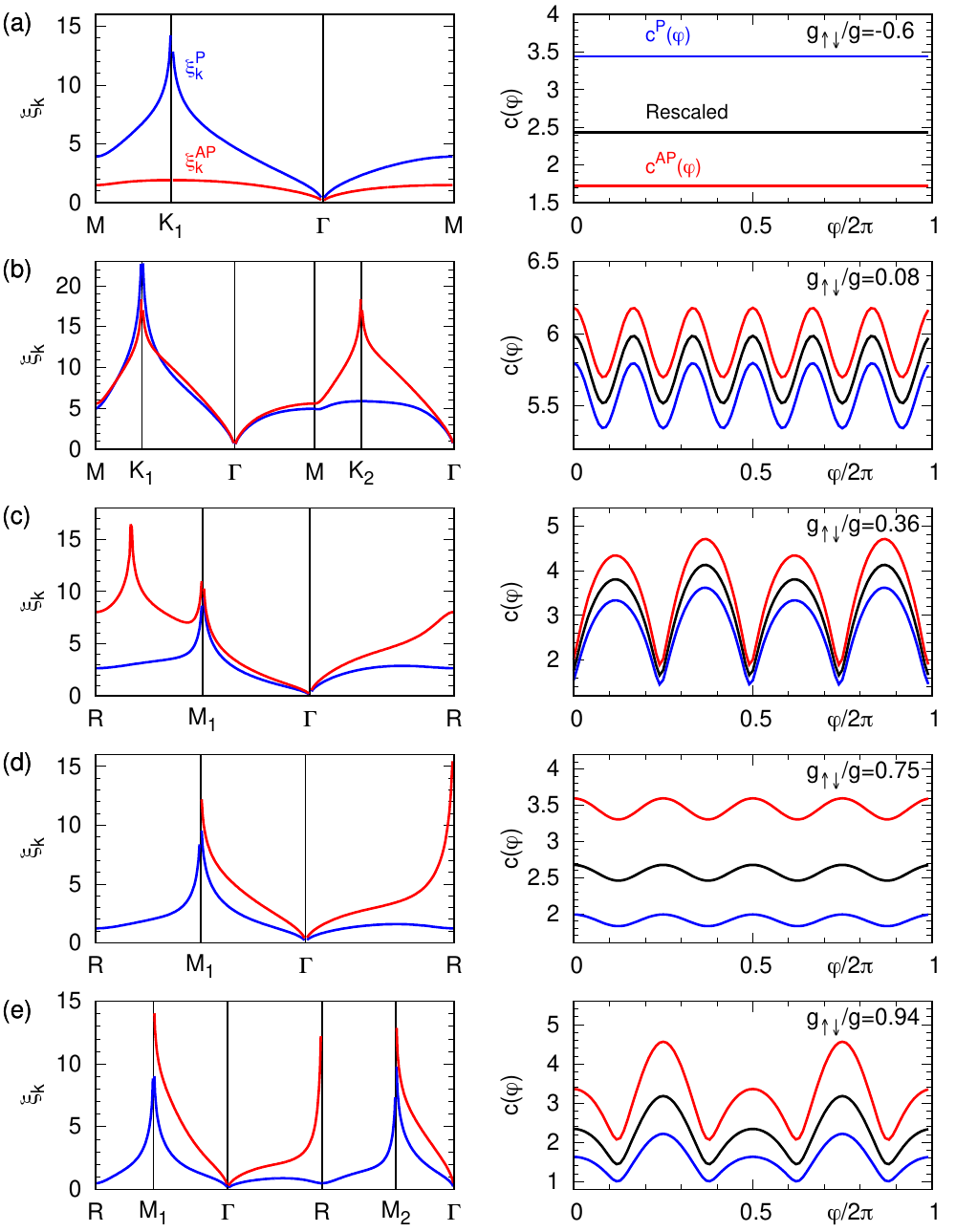}%{nu20_N97_ES_MF.pdf}
\caption{\label{fig:ES_GP}
Left panels: single-particle ES $\xi_\kv^{\mathrm{P}/\mathrm{AP}}$ for parallel (P; blue) and antiparallel (AP; red) fields 
for the same cases (a)-(e) as in Figs.\ \ref{fig:spec_unresc}, \ref{fig:spec_resc}, and \ref{fig:aniso_energy}. 
Calculations are based on Eq.\ \eqref{eq:BoseDistributionFunction} and 
done along the paths indicated by dotted arrows shown in the lower panels of Fig.\ \ref{fig:Phases_GP}. 
The spectra $\xi_\kv$ show divergence at some high-symmetry points and along lines in the first Brillouin zone. 
In (d) and (e), in particular, divergence occurs along the edges of the Brillouin zone for antiparallel fields and thus the value of $\xi_\kv$ is not shown there. 
Right panels: dimensionless functions $c^{\mathrm{P}/\mathrm{AP}}(\varphi)$ 
that express the anisotropy of the ES around the $\Gamma$ point as in Eq.\ \eqref{eq:xi_sqrt_k_vorlat}. 
These are calculated from $\xi_\kv^{\mathrm{P}/\mathrm{AP}}$ 
along a circular path $\kv=k(\cos\varphi,\sin\varphi)$ with $k=0.001a/\ell^2$ and $\varphi \in [0,2\pi)$. 
With proper rescaling, the curves for parallel and antiparallel fields coincide perfectly up to numerical precision, 
confirming the rescaling relation in Eq.\ \eqref{eq:c_phi_rescale}. 
Namely, $(\gb_+/\gb_-)^{1/4}c^\mathrm{P}(\varphi)$ and $(\gb_-/\gb_+)^{1/4}c^\mathrm{AP}(\varphi)$ share the same curves shown in black. 
%
%Single-particle ES $\xi_\kv$ along the path that connect the high-symmetry points of the first Brillouin zone (left panels) and the angle-dependence of $\xi_\kv/(k\ell)^{1/2}$ around the $\Gamma$ point (right panels) for (a) overlapping triangular, (b) interlaced triangular, (c) rhombic, (d) square and (e) rectangular lattices for the parallel (blue) and antiparallel (red) fields with $\nu=\infty$.  
% In the right panels, the calculations are performed along the circular path centered around the $\Gamma$ point with the radius $0.001a/\ell^2$. 
% The dotted lines indicate the analytical expression obtained by the effective field theory; here, we use the values of the elastic constants obtained in \cite{Yoshino19}.
% The spectra $\xi_\kv$ show divergences at some high-symmetry points or along lines in the first Brillouin zone. 
% In particular, in (d) and (e) the divergence occurs along the edges of the Brillouin zone for antiparallel fields and thus the value of $\xi_\kv$ is not shown there. 
% These divergences should correspond to the decoupling of modes into the spin-$\ua$ and the spin-$\da$ components.  
}
\end{figure*}
%############################

% [ Intercomponent entanglement ]--------------------------------------
Here we present numerical results on the intercomponent ES and EE in the Bogoliubov ground state. 
Calculations are based on the formulation in Sec.\ \ref{sec:BogoliubovTheory_ent}. 
The obtained results are compared with the field-theoretical results in Sec.\ \ref{sec:ent_vorlat}.

% [ Entanglement spectrum: rescaling relation ]--------------------------------------
The left panels of Fig.\ \ref{fig:ES_GP} display the single-particle ES $\xi_\kv$ 
for the same cases as in Figs.\ \ref{fig:spec_unresc}, \ref{fig:spec_resc}, and \ref{fig:aniso_energy}. 
Around the $\Gamma$ point, $\xi_\kv$ can be fit well by the square-root dispersion relation 
\begin{equation}\label{eq:xi_sqrt_k_vorlat}
 \xi_\kv = c(\varphi)\sqrt{k\ell}
\end{equation}
for $\kv=k(\cos\varphi,\sin\varphi)$, where $c(\varphi)$ is a dimensionless function that expresses the anisotropy. 
The right panels of Fig.\ \ref{fig:ES_GP} show the function $c(\varphi)$ determined from the data of $\xi_\kv$ 
along a circular path around the $\Gamma$ point. 
With proper rescaling, the curves for parallel and antiparallel fields are found to coincide perfectly up to numerical precision; 
furthermore, the rescaled curves are found to agree accurately with $4[C(\varphi)/D(\varphi)]^{1/2}$ (not shown), 
where $\sqrt{C(\varphi)}$ and $\sqrt{D(\varphi)}$ are shown in Fig.\ \ref{fig:aniso_energy}. 
Thus, we can confirm the rescaling relation for the entanglement spectra in Eq.\ \eqref{eq:c_phi_rescale}.

% [ Entanglement spectrum: divergences ]--------------------------------------
In the left panels of Fig.\ \ref{fig:ES_GP}, we also find that $\xi_\kv$ diverges at some high-symmetry points and along lines in the Brillouin zone. 
For square (d) and rectangular (e) lattices in antiparallel fields, in particular, divergence occurs along the edges of the Brillouin zone 
(thus $\xi_\kv$ is not shown along the paths $R\to M_1$ and $R\to M_2$). 
In our previous work \cite{Yoshino19} (see Appendix D therein), it has been found that 
at the $M_1$ and $M_2$ points for rhombic, square and rectangular lattices and for both parallel and antiparallel fields, 
the Bogoliubov Hamiltonian matrix $\Mcal(\kv)$ has the structure in which the spin-$\ua$ and $\da$ components are decoupled. 
Therefore, $\xi_\kv$ naturally diverges at these points. 
However, we have not been able to find such a simple structure of $\Mcal(\kv)$ along the edges of the Brillouin zone for square (d) and rectangular (e) lattices in antiparallel fields. 

%{\bf (The following consideration should be written after we examine the Bogoliubov matrix element in more detial.)}
%The divergence of $\xi_\kv$ along the whole edges of the Brillouin zone for antiparallel fields is thus an unexpected feature. 
%In fact the Bogoliubov matrix of the $M$ point for the rhombic, square and rectangle lattices for both the cases of parallel and antiparallel fields, where the entanglement spectrum diverges, indicates that the spin-$\ua$ and $\da$ components are decoupled, 
%while the Bogoliubov matrix of the $K_1$ point for the overlapping and interlaced triangular lattices in antiparallel fields, where the entanglement spectrum converges, indicates that two-components are not decoupled although the Bogoliubov excitation spectra degenerate for all the high symmetry points \cite{Yoshino19}. 
%In contrast the divergence of the entanglement spectrum indicates that the spin-$\ua$ and $\da$ components are decoupled. 
%We do not have any explanation for the correspondence of the nodal lines and points for the parallel (antiparallel) magnetic fields and the divergence of the entanglement spectrum for the antiparallel (parallel) magnetic fields. 

%############################
\begin{figure}
\centering\includegraphics[width=16cm , angle=0]{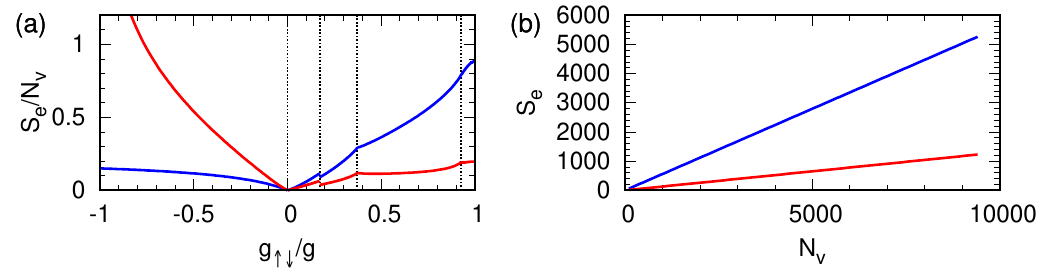}%{nu20_EE_SQU75d-2_Nv_N10_all.pdf}
%\centering\includegraphics[width=8cm , angle=0]{nu20_EE_SQU75e-2-EEsubEE_N10.pdf}
\caption{\label{fig:EE_vorlat} 
(a) Intercomponent EE per flux quantum, $S_\erm/\Nvor$, as a function of $g_{\ua\da}/g$ for parallel (blue) and antiparallel (red) fields for $\Nvor=97^2$. 
This is calculated from numerically obtained $\xi_\kv$ using Eq.\ \eqref{eq:EE}. 
Vertical dashed lines indicate the transition points. 
%In consistency with the field-theoretical result in Sec.\ \ref{sec:ent_vorlat} [see, in particular, Eq.\ \ref{eq:Se_vorlat} therein], 
%$S_\erm/\Nvor$ is found to be larger for parallel (antiparallel) fields when the intercomponent interaction $g_{\ua\da}$ is repulsive (attractive). 
(b) Intercomponent EE $S_\erm$ versus $\Nvor$ for $g_{\ua\da}/g=0.75$. 
A fit to the form $S_\erm=\alpha_1 \Nvor - \alpha_2\ln \Nvor+\alpha_3$ gives 
$(\alpha_1,\alpha_2,\alpha_3)=(0.560,0.236,-0.70)$ for parallel fields and 
$(\alpha_1,\alpha_2,\alpha_3)=(0.130,0.243,-0.06)$ for antiparallel fields. 
%$(\alpha_1,\alpha_2,\alpha_3)=(0.559702,0.235899,-0.701552)$ for parallel fields and 
%$(\alpha_1,\alpha_2,\alpha_3)=(0.130402,0.243383,-0.0583632)$ for antiparallel fields. 
%(c) $S_\erm-\alpha_1 \Nvor$ versus $\Nvor$.  A logarithmic scale is taken for the horizontal axis. 
%The presence of the subleading logarithmic term can clearly be seen in this figure. 
%{\bf Notes: $S_\erm-\alpha_1 \Nvor$ for the vertical axis label in (c). Data points and fit lines should be shown for (b) and (c).}
}
\end{figure}
%############################

% [ Entanglement entropy ]--------------------------------------
Figure \ref{fig:EE_vorlat}(a) shows the intercomponent EE per flux quantum   
as a function of the ratio $g_{\ua\da}/g$ for parallel (blue) and antiparallel (red) fields. 
We find that for repulsive (attractive) $g_{\uparrow\downarrow}$, the EE tends to be larger for parallel (antiparallel) fields, 
in consistency with the field-theoretical result in Sec.\ \ref{sec:ent_vorlat} [see Eq.\ \eqref{eq:Se_vorlat}].
This behavior is in qualitative agreement with the exact diagonalization results in a quantum (spin) Hall regime 
with $\nu=\Ocal(1)$ \cite{Furukawa14,Furukawa17}, 
in which product states of nearly uncorrelated quantum Hall states are found to be robust 
for an intercomponent attraction (repulsion) in the case of parallel (antiparallel) fields. 

In Figs.\ \ref{fig:EE_vorlat}(b), % and (c), 
we examine the scaling of the intercomponent EE $S_\erm$ as a function of $\Nvor$. 
As seen in %Fig.\ \ref{fig:EE_vorlat}(b), 
this figure, 
the dominant part of the scaling is given by a volume-law behavior $\alpha_1 \Nvor$; 
such a volume-law contribution is standard for an extensive cut as discussed here, and has also been found elsewhere 
\cite{Lundgren13,Chen13,Furukawa11,Xu11,Mollabashi14}. 
In agreement with the field-theoretical result in Eq.\ \eqref{eq:Se_Nv_logNv}, 
we find that the data are well fitted by the form $S_\erm=\alpha_1 \Nvor - \alpha_2\ln \Nvor+\alpha_3$ 
and that the coefficient $\alpha_2$ obtained by the fitting is close to $1/4$. 
%The presence of the subleading logarithmic contribution can clearly be seen Fig.\ \ref{fig:EE_vorlat}(c),
%in which the leading volume-law contribution $\alpha_1 \Nvor$ is subtracted from $S_\erm$. 

%************************************************
\subsection{Fraction of depletion}\label{sec:numerics_dep}
%************************************************

%############################
\begin{figure*}
\centering
\includegraphics[width=16cm , angle=0]{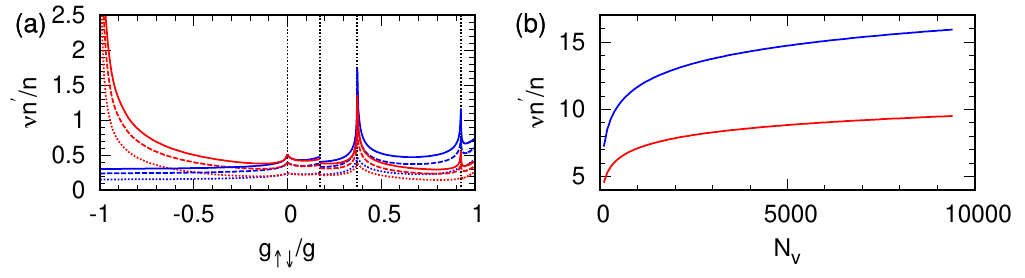}%{nu20_Qdep_SQUNv_N10.pdf}
\caption{\label{fig:dep_vorlat} 
(a) Fraction of depletion $n'/n$ (scaled by $\nu^{-1}$) as a function of $g_{\ua\da}/g$ for parallel (blue) and antiparallel (red) fields 
for $\Nvor=5^2$ (dotted), $11^2$ (dashed), and $19^2$ (solid). 
The calculations are based on Eq.\ \eqref{eq:depletion_Vcal}. 
Vertical dashed lines indicate the transition points. 
%In consistency with the field-theoretical result in Eq.\ \eqref{eq:dep_log}, $\nu n'/n$ shows a logarithmic increase as a function of $\Nvor$, 
%as seen in its equidistant increase among the three values of $\Nvor$. 
%Furthermore, $\nu n'/n$ is larger for parallel (antiparallel) fields when the intercomponent interaction $g_{\ua\da}$ is repulsive (attractive). 
(b) $\nu n'/n$ versus $\Nvor$ for $g_{\ua\da}/g=0.75$. 
A fit to the form $\nu n'/n=\gamma_1 \ln \Nvor+\gamma_2$ gives
$(\gamma_1,\gamma_2)=(1.91,-1.53)$ for parallel fields and $(\gamma_1,\gamma_2)=(1.09,-0.40)$ for antiparallel fields. 
%$(\gamma_1,\gamma_2)=(1.91026,-1.52925)$ for parallel fields and $(\gamma_1,\gamma_2)=(1.08532,-0.397055)$ for antiparallel fields. 
%{\bf Notes: In (b), data points and fit lines should be shown. Logarithmic scale for the horizontal axis.}
}
\end{figure*}
%############################

Figure \ref{fig:dep_vorlat} shows numerical results on the fraction of depletion $n'/n$ (scaled by $\nu^{-1}$). 
As seen in Fig.\ \ref{fig:dep_vorlat}(a), for an intercomponent repulsion (attraction), this quantity tends to be larger for parallel (antiparallel) fields, 
indicating stronger quantum fluctuations. 
This is in agreement with the field-theoretical result in Eq.\ \eqref{eq:dep_log}. 
At the transition point between interlaced triangular and rhombic lattices, 
the fraction of depletion changes discontinuously owing to a discontinuous change in the lattice structure. 
Meanwhile, the fraction of depletion diverges at both the transition points between rhombic, square, and rectangular lattices; 
this seems to be related to rapid changes in the inner angle and the aspect ratio in the mean-field ground state 
as shown in Fig.\ \ref{fig:parameters}. 
In Fig.\ \ref{fig:dep_vorlat}(b), we examine the scaling of $\nu n'/n$ as a function of $\Nvor$. 
The data are well fitted by the logarithmic form $\nu n'/n=\gamma_1 \ln \Nvor+\gamma_2$, 
in agreement with the field-theoretical result in Eq.\ \eqref{eq:dep_log}. 

%************************************************
\subsection{Ground-state phase diagrams}\label{sec:phase_diagrams}
%************************************************

%############################
\begin{figure*}
 \centering
  \includegraphics[width=16cm, angle=0]{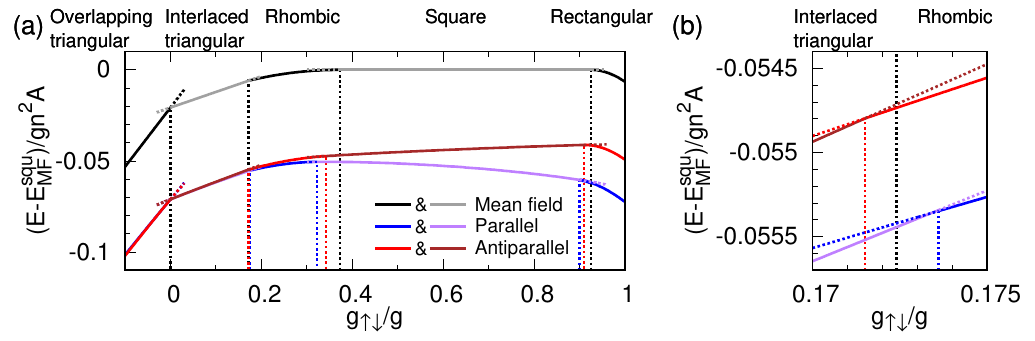}%{nu20_EGS_N97.pdf}
 \caption{\label{fig:GS energy}  
 (a) Ground-state energy [Eq.\ \eqref{eq:Egs/atom}] as a function of $g_{\ua\da}/g$.  
 The mean-field energy (black and grey) is appreciably changed by a quantum correction, as shown for both parallel (blue and purple) and antiparallel (red and brown) fields for $\nu=20$. 
 The mean-field energy for the square lattices, 
 $E^\mathrm{squ}_\mathrm{MF}/(gn^2A)=1.18034+0.834627 g_{\ua\da}/g$, 
 is subtracted to emphasize the changes due to quantum fluctuations. 
 Vertical dashed lines indicate the transition points, and alternating colors correspond to different phases. 
 In particular, the transition points between rhombic-, square- and rectangular lattices shift appreciably owing to quantum corrections. 
 In contrast, the transition point $g_{\ua\da}/g=0$ between overlapping and interlaced triangular lattices remain unchanged by quantum corrections. 
 (b) Enlarged view of (a) around the transition point between interlaced triangular and rhombic lattices, which also shows small shifts due to quantum corrections.  
% Top: Energy of a ground state within the mean-field theory (broken lines and crosses) and the real part of a ground state energy for which we take the LHY correction into account (filled dots). We indicate transition points within the mean-field theory by vertical lines. Quantum fluctuation corrects the ground state energy within the mean field theory (grey and black) for the parallel fields case (blue and purple) and the antiparallel fields case (red and brown).
}
\end{figure*}
%############################

% [ Declaration, Assumption & detail of minimization ]--------------------------------------
Here we analyze how quantum fluctuations affect the ground-state phase diagrams for parallel and antiparallel fields. 
%Within the GP mean-field theory, the ground-state phase diagram has been obtained by minimizing the mean-field energy [the first term of Eq.\ \eqref{eq:Egs/atom}] 
%with respect to the inner angle $\theta$, the ratio $b/a$, and the displacement parameters $u_1$ and $u_2$ with the magnetic length $\ell$ held fixed \cite{Mueller02}. 
%This mean-field analysis has led to the five types of lattice structures that depend on the ratio $g_{\ua\da}/g$ as shown in Fig.\ \ref{fig:Phases_GP}. 
%Below we assume that the same types of structures appear in the presence of quantum fluctuations as well, 
%and examine the correction to the ground-state energy due to zero-point fluctuations [the term proportional to $\nu^{-1}$ in Eq.\ \eqref{eq:Egs/atom}]. 
The GP mean-field analyses \cite{Mueller02,Kasamatsu03,Kasamatsu05,Mingarelli18} 
have led to the five types of lattice structures that depend on $g_{\ua\da}/g$ as shown in Fig.\ \ref{fig:Phases_GP}. 
We assume that the same types of structures appear in the presence of quantum fluctuations as well, 
and examine a quantum correction to the ground-state energy. 

%############################
\begin{figure*}
 \centering
  \includegraphics[width=12cm, angle=0]{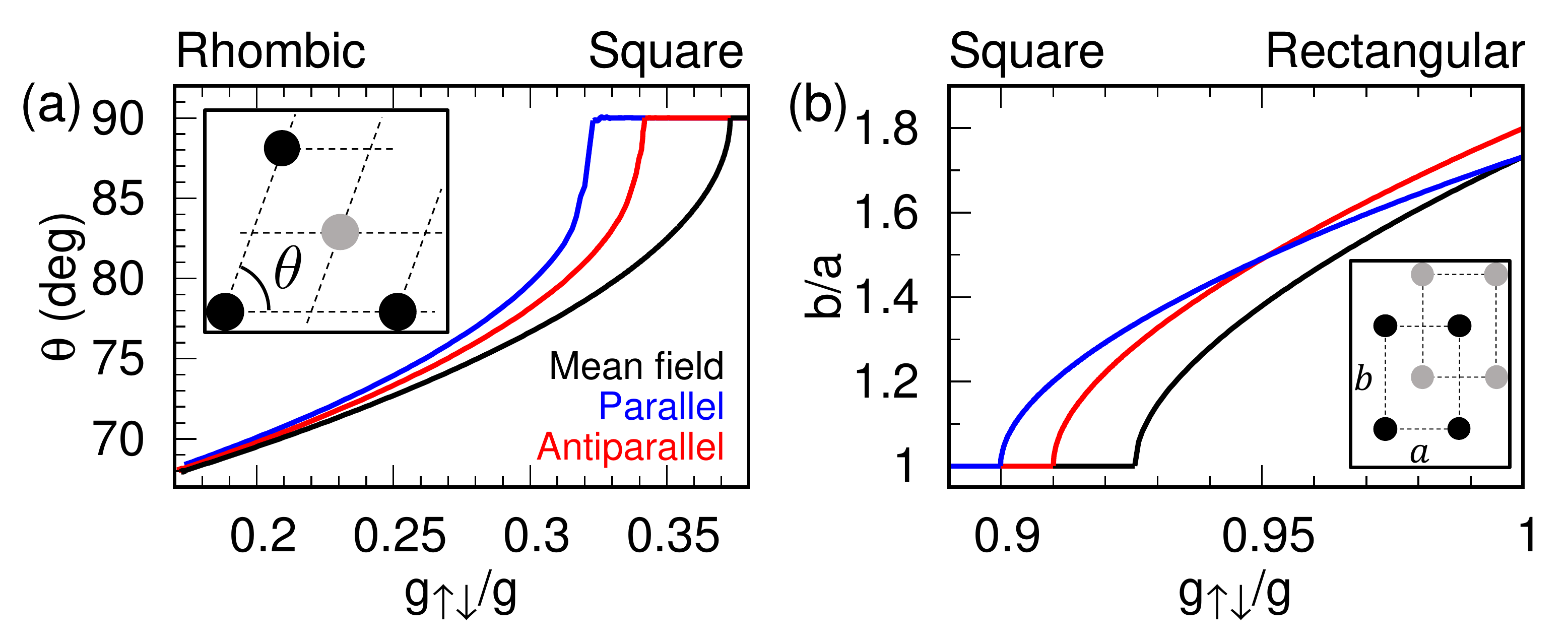}%{nu20_parmt_N97_all.pdf}
 \caption{\label{fig:parameters}  
 (a) The inner angle $\theta$ of rhombic lattices and (b) the aspect ratio $b/a$ of rectangular lattices, plotted against $g_{\ua\da}/g$. 
 The mean-field results (black) \cite{Mueller02} are changed appreciably by quantum corrections, as shown both for parallel (blue) and antiparallel (red) fields. 
 Insets show the definitions of $\theta$ and $b/a$. 
 }
\end{figure*}
%############################

% [ Ground state energy and parameters ]--------------------------------------
Figure\ \ref{fig:GS energy} shows the mean-field ground-state energy 
as well as those with quantum corrections for parallel and antiparallel fields, 
where the filling factor is $\nu=20$.  
Here, the energies for rhombic and rectangular lattices are minimized with respect to the inner angle $\theta$ and the aspect ratio $b/a$, respectively.
As seen in Fig.\ \ref{fig:GS energy} (a), the transition points between rhombic, square and rectangular lattices shift appreciably owing to quantum corrections; 
the shift occurs more significantly for parallel fields. 
In contrast, the transition point $g_{\ua\da}/g=0$ between overlapping and interlaced triangular lattices remains unchanged by quantum corrections. 
While the shift of the transition point between interlaced triangular and rhombic lattices is not clearly seen in Fig\ \ref{fig:GS energy}(a), 
the shift indeed occurs as seen in the enlarged plot in Fig.\ \ref{fig:GS energy}(b). 

Figure \ref{fig:parameters}(a) and (b) show the inner angle $\theta$ of rhombic lattices and the aspect ratio $b/a$ of rectangular lattices, respectively, 
which are obtained through the one-parameter minimization of the ground-state energy \eqref{eq:Egs/atom}. 
In the mean-field results, both the inner angle $\theta$ and the aspect ratio $b/a$ show rapid changes near the transition points to square lattices \cite{Yoshino19}.
In these regimes, the systems are expected to be highly susceptible to quantum fluctuations. 
Indeed, changes in $\theta$ and $b/a$ are enhanced in these regimes, 
which explains the large shifts of the transition points as demonstrated in Fig.\ \ref{fig:GS energy}(a). 

%Indeed, the fraction of depletion diverges at the transition points, as seen in Fig.\ \ref{fig:dep_vorlat}. 
%This explains why the changes in $\theta$ and $b/a$ are enhanced around the transition points 
%and why the shift of the transition point is larger for parallel fields. 

%%%%%%%%%%%%%%%%%%%%%%%%%%%%%%%%%%%%%%%%%%%%%%%%%
\section{Summary and outlook}\label{sec:Summary}
%%%%%%%%%%%%%%%%%%%%%%%%%%%%%%%%%%%%%%%%%%%%%%%%%

In this paper, we have presented a detailed comparative study of vortex lattices of binary BECs in parallel and antiparallel synthetic magnetic fields. 
Within the GP mean-field theory valid for high filling factors $\nu\gg 1$, the two types of fields are known to lead to the same phase diagram 
that consists of a variety of vortex lattices \cite{Mueller02,Kasamatsu03,Kasamatsu05,Mingarelli18,Furukawa14}. 
We have formulated an improved effective field theory for such vortex lattices 
by introducing renormalized coupling constants for coarse-grained densities, 
and studied properties of collective modes and ground-state intercomponent entanglement. 
We have also performed numerical calculations based on the Bogoliubov theory with the LLL approximation to confirm the field-theoretical predictions. 
We have shown that the low-energy excitation spectra for the two types of fields are related to each other by suitable rescaling using the renormalized coupling constants
[Eqs.\ \eqref{eq:E12_rescaling} and \eqref{eq:energy_rescaling} and Figs.\ \ref{fig:spec_resc} and \ref{fig:aniso_energy}]. 
By calculating the intercomponent EE in the ground state, we have found that for an intercomponent repulsion (attraction), 
the two components are more strongly entangled in the case of parallel (antiparallel) fields [Eq.\ \eqref{eq:Se_vorlat} and Fig.\ \ref{fig:EE_vorlat}(a)],  
in qualitative agreement with recent numerical results for a quantum (spin) Hall regime \cite{Furukawa14,Furukawa17}. 
As a by-product, we have also found that the ES exhibits an anomalous square-root dispersion relation [Eq.\ \eqref{eq:xi_F_G_vorlat} and Fig.\ \ref{fig:ES_GP}], 
and that the EE exhibits a volume-law scaling followed by a subleading logarithmic term [Eq.\ \eqref{eq:Se_Nv_logNv} and Fig.\ \ref{fig:EE_vorlat}(b)]. 
Finally, we have investigated the effects of quantum fluctuations on the phase diagrams 
by calculating the correction to the ground-state energy due to zero-point fluctuations in the Bogoliubov theory [Eq.\ \eqref{eq:Egs/atom} and Fig.\ \ref{fig:GS energy}]. 
We have found that the boundaries between rhombic-, square-, and rectangular-lattice phases shift appreciably with a decrease in $\nu$. 

% [ Outlook ]--------------------------------------
We have seen a similarity between the regimes of high ($\nu\gg 1$) and low ($\nu=O(1)$) filling factors
in the behavior of intercomponent entanglement. 
It will be interesting to investigate how the two regimes are connected 
by applying sophisticated numerical methods such as a variational wave function \cite{Kwasigroch12} 
and the infinite density matrix renormalization group \cite{Geraedts17}.   
Furthermore, the similarity between the two regimes suggests that 
the behavior of intercomponent entanglement does not depend on the details of the systems 
and can be universal for a wide range of Hamiltonians. 
In fact, it has been found in lattice models that two coupled bosonic Laughlin states with opposite chiralities 
(i.e., fractional quantum spin Hall states \cite{Bernevig06}) 
are more robust against an intercomponent repulsion than the ones with the same chirality \cite{Repellin14}. 
The stability of fractional quantum spin Hall states against an intercomponent repulsion 
has also been discussed in fermionic models \cite{Neupert11,LiW14,ChenH12,Ghaemi12}. 
Comparative investigation of multicomponent systems in gauge fields as conducted in the present work 
will be a useful approach for exploring universal features of interacting topological states of matter. 

%bilayer quantum Hall systems \cite{Girvin97} and quantum spin Hall systems \cite{Bernevig06}. 
%Comparative studies of multicomponent systems with the same and opposite chiralities 
%will lead to a unified understanding 

%The ground-state phase diagram can be studied in detail by minimizing the corrected ground state energy 
%with all parameters (the inner angle, the aspect ratio and the vortex displacement between two components).
%The corrected ground state energy taking third or higher order of the fluctuation into account can be investigated 
%by using the trial wave function introduced in a scalar BEC in \cite{Kwasigroch12}.
%For binary BECs with vortex lattices where two components are coupled by density-density interaction, 
%we have shown the intriguing square-root entanglement spectrum arise.
%By investigating the entanglement spectrum of various systems with a coupling of two components, 
%the ES with characteristic fractional power may be obtained. 
%
%We infer that different behavior in the energy dispersion relations in the symmetric and antisymmetric channels 
%is the key to the anomalous dispersion relation in the ES. 

%%%%%%%%%%%%%%%%%%%%%%%%%%%%%%%%%%%%%%%%%%%%%%%%%
\section*{Acknowledgments}
%%%%%%%%%%%%%%%%%%%%%%%%%%%%%%%%%%%%%%%%%%%%%%%%%

% [ Acknowledgments ]--------------------------------------
TY is grateful to Kazuya Fujimoto, Yuki Kawaguchi, and Terumichi Ohashi for stimulating discussions during his stay at Nagoya University. 
This work was supported by KAKENHI Grant No.\ JP18H01145 and No.\ JP18K03446, 
a Grant-in-Aid for Scientific Research on Innovative Areas ``Topological Materials Science'' (KAKENHI Grant No.\ JP15H05855) 
from the Japan Society for the Promotion of Science (JSPS), and Keio Gijuku Academic Development Funds.
TY was supported by JSPS through the Program for Leading Graduate School (ALPS). 

\appendix

%%%%%%%%%%%%%%%%%%%%%%%%%%%%%%%%%%%%%%%%%%%%%%%%%
\section{Entanglement spectrum and entanglement Hamiltonian}\label{app:entanglement}
%%%%%%%%%%%%%%%%%%%%%%%%%%%%%%%%%%%%%%%%%%%%%%%%%

Here we describe details of the calculation of the ES $\xi_\kv$ and the coefficients $F_\kv$ and $G_\kv$ in the entanglement Hamiltonian in Sec.\ \ref{sec:ent_vorlat}.

% [ Correlations in the oscillator ground state ]--------------------------------------
For preparation, we calculate the phase and density correlators in the oscillator ground state $\ket{0^\osc}$ of the total system. 
By using Eqs.\ \eqref{eq:CanoTrnf_Scaling}, \eqref{eq:CanoTrnf_Diagonalization}, and \eqref{eq:CanoTrnf_VortexLatticeBogolon}, 
the phase and the density of the spin-$\ua$ component are expressed in terms of the bogolon operators as
\begin{subequations}\label{eq:theta_n_gamma_vorlat}
\begin{align}
 \theta_{\kv,\ua} 
 &=\frac12 \left( r_\kv, r_\kv^{-1} \right) U(\kv) \begin{pmatrix} \bar{\theta}_{\kv,1} \\ \bar{\theta}_{\kv,2} \end{pmatrix}
 = \frac12\sum_{j=1,2}R_{\kv,j}\bar{\theta}_{\kv,j} 
 =\frac{1}{2\sqrt{2n}} \sum_{j=1,2} R_{\kv,j}\zeta_{\kv,j}^{-1} \left(\gamma_{\kv,j}+\gamma_{-\kv,j}^\dagger \right),\\
 n_{\kv,\ua}
 &=\left( r_\kv^{-1}, r_\kv \right) U(\kv) \begin{pmatrix} \bar{n}_{\kv,1} \\ \bar{n}_{\kv,2} \end{pmatrix}
 =\sum_{j=1,2} R_{\kv,3-j} \bar{n}_{\kv,j} 
 = \frac{1}{\irm} \sqrt{\frac{n}{2}} \sum_{j=1,2} R_{\kv,3-j} \zeta_{\kv,j} \left(\gamma_{\kv,j}-\gamma_{-\kv,j}^\dagger \right).
\end{align}
\end{subequations}
Here we introduce 
\begin{equation}
 R_{\kv,1}:=r_\kv       \cos\frac{\chi_\kv}{2}+\irm r_\kv^{-1}\sin\frac{\chi_\kv}{2} ,~~~
 R_{\kv,2}:=r_\kv^{-1}\cos\frac{\chi_\kv}{2}+\irm r_\kv       \sin\frac{\chi_\kv}{2},
\end{equation}
which satisfy $R_{-\kv,j}=R_{\kv,j}^\ast~(j=1,2)$. 
As $\ket{0^\osc}$ is the vacuum for bogolons, i.e., $\gamma_{\kv,\pm}\ket{0^\osc}=0$ for all $\kv\ne\zerov$, 
the correlators of the operators in Eq.\ \eqref{eq:theta_n_gamma_vorlat} are calculated as
\begin{subequations}\label{eq:CorrFunc_VL}
\begin{align}
 &\bra{0^\osc}\theta_{-\kv,\ua}\theta_{\kv,\ua}\ket{0^\osc}  =\frac{1}{8n} \sum_{j=1,2} |R_{\kv,j}|^2 \zeta_{\kv,j}^{-2}
 =\frac{1}{8n\zeta_{\kv,1}\zeta_{\kv,2}} \left(|R_{\kv,1}|^2\frac{\zeta_{\kv,2}}{\zeta_{\kv,1}} +|R_{\kv,2}|^2\frac{\zeta_{\kv,1}}{\zeta_{\kv,2}}\right),\\
 &\bra{0^\osc}n_{-\kv,\ua}n_{ \kv,\ua}\ket{0^\osc}  =\frac{n}{2} \sum_{j=1,2} |R_{\kv,3-j}|^2 \zeta_{\kv,j}^2
 =\frac{n\zeta_{\kv,1}\zeta_{\kv,2}}{2} \left(|R_{\kv,1}|^2\frac{\zeta_{\kv,2}}{\zeta_{\kv,1}} +|R_{\kv,2}|^2\frac{\zeta_{\kv,1}}{\zeta_{\kv,2}}\right).
\end{align}
\end{subequations}

% [ Correlations in the oscillator ground state ]--------------------------------------
We now require that the correlators \eqref{eq:Correlator_FkGk} obtained from the Gaussian ansatz \eqref{eq:Entanglement_Ham_oscillation} 
be equal to the ones \eqref{eq:CorrFunc_VL} obtained for the oscillator ground state $\ket{0^\mathrm{osc}}$.
We then find
\begin{subequations}\label{eq:f_FG_vorlat}
\begin{align}
 f_\Brm (\xi_\kv) + \frac12
 &=\sqrt{ \bra{0^\osc}\theta_{-\kv,\ua}\theta_{\kv,\ua}\ket{0^\osc}
     \bra{0^\osc}n_{-\kv,\ua}n_{\kv,\ua}\ket{0^\osc}}   
 =\frac14\left(|R_{\kv,1}|^2\frac{\zeta_{\kv,2}}{\zeta_{\kv,1}}
  +|R_{\kv,2}|^2\frac{\zeta_{\kv,1}}{\zeta_{\kv,2}}\right), \\
% \label{eq:BoseDistributionFunc_correlator} \\
 \sqrt{\frac{F_\kv}{G_\kv}} 
 &=\frac{1}{n}\sqrt{\frac{\bra{0^\osc}n_{-\kv,\ua}n_{\kv,\ua}\ket{0^\osc}}
     {\bra{0^\osc}\theta_{-\kv,\ua}\theta_{\kv,\ua}\ket{0^\osc}} }
 =2\zeta_{\kv,1}\zeta_{\kv,2}.
% \label{eq:EntanglementHam_s_k} 
\end{align}
\end{subequations}
In the long-wavelength limit $k\ell \ll 1$, %where $\tilde{\Gamma}_-(\kv) \gg |\Gamma(\kv)| \gg |\tilde{\Gamma}_+(\kv)|$, 
we have
\begin{equation}\label{eq:zeta_R_vorlat}
\begin{split}
 &\zeta_{\kv,1}
 %\left[\frac{\Gamma_-(\kv)}{2g_\pm n^2}\right]^{1/4}
 \approx\left[\frac{g D(\varphi)}{2\gb_\pm}\right]^{1/4} \sqrt{k\ell} , ~~
 \zeta_{\kv, 2}
 %\left[\frac{\Gamma_-(\kv)}{2g_\pm n^2}\right]^{1/4}
 \approx \left[\frac{g C(\varphi)}{2\gb_\mp}\right]^{1/4}k\ell , ~~
 \cos\chi_\kv \approx \mp 1,\\
 &|R_{\kv,1}|^2 =r_\kv^2 \frac{1+\cos\chi_\kv}2 + r_\kv^{-2}\frac{1-\cos\chi_\kv}{2}\approx r_\zerov^{\mp 2} = \sqrt{\frac{\gb_\mp}{\gb_\pm}},\\
 &|R_{\kv,2}|^2 =r_\kv^{-2} \frac{1+\cos\chi_\kv}2 + r_\kv^2\frac{1-\cos\chi_\kv}{2}\approx r_\zerov^{\pm 2} = \sqrt{\frac{\gb_\pm}{\gb_\mp}},
\end{split}
\end{equation}
where Eq.\ \eqref{eq:m_k} is used. Then, $f_\Brm(\xi_\kv)$ and $\sqrt{F_\kv/G_\kv}$ in Eq.\ \eqref{eq:f_FG_vorlat} are calculated as
\begin{equation}\label{eq:BoseDist._VorLat} 
 f_\Brm(\xi_\kv)
 %\approx \frac14 |R_{\kv,2}|^2 \frac{\zeta_{\kv,1}}{\zeta_{\kv,2}}
 \approx \frac14\left[\frac{\gb_\pm D(\varphi)}{\gb_\mp C(\varphi)}\right]^{1/4} (k\ell)^{-1/2} , ~~~
 \sqrt{\frac{F_\kv}{G_\kv}}
 \approx \left[ \frac{4g^2C(\varphi)D(\varphi)}{\gb_+ \gb_-} \right]^{1/4}(k\ell)^{3/2}. 
\end{equation}
from which we obtain Eq.\ \eqref{eq:xi_F_G_vorlat}. 

%%%%%%%%%%%%%%%%%%%%%%%%%%%%%%%%%%%%%%%%%%%%%%%%
\section{Phase correlation function}\label{app:phase_corr}
%%%%%%%%%%%%%%%%%%%%%%%%%%%%%%%%%%%%%%%%%%%%%%%%

Here we describe the derivation of the phase correlation function \eqref{eq:corr_theta_r0_vorlat}. 
Using Eq.\ \eqref{eq:corr_theta_n_vorlat_FG}, this correlation function is expressed as
\begin{equation}\label{eq:corr_theta_G}
 \langle \left[ \theta_\ua(\rv)-\theta_\ua(\zerov) \right]^2\rangle  \approx  \frac{2}{nF\ell^2} \left[ G(\zerov;\alpha)-G(\rv;\alpha)\right].
\end{equation}
Here, $G(\rv;\alpha)$ is Green's function for the 2D Poisson equation 
\begin{equation}\label{eq:Green_Poisson}
 G(\rv;\alpha) 
 = \frac{1}{A} \sum_{\kv\ne\zerov} \frac{1}{k^2} \erm^{-\alpha k+\irm \kv\cdot\rv} 
 = \frac{1}{A} \sum_{\kv\ne\zerov} \frac{1}{k^2} \erm^{-\alpha k+\irm kr\cos\theta}, 
\end{equation}
where we express the wave vector $\kv$ using the polar coordinate $(k,\theta)$ with $\theta$ being the angle relative to $\rv$. 
Although the logarithmic behavior of Green's function for the 2D Poisson equation is known, 
we derive it within the present regularization scheme using the convergence factor $\erm^{-\alpha k}$. 

By differentiating Eq.\ \eqref{eq:Green_Poisson} with respect to $r$, we have
\begin{equation}
\begin{split}
 -\frac{\partial}{\partial r} G(\rv;\alpha) 
 &= \frac{1}{A} \sum_{\kv\ne\zerov} \frac{-\irm\cos\theta}{k} \erm^{-\alpha k+\irm kr\cos\theta} 
 = \frac{1}{(2\pi)^2} \int_0^{2\pi}\!\!\!\! \drm\theta \int_0^\infty\!\!\!\! \drm k ~(-\irm\cos\theta) \erm^{-k(\alpha -\irm r\cos\theta)} \\
 &= \frac{1}{(2\pi)^2} \int_0^{2\pi} \drm\theta \frac{-\irm \cos\theta}{\alpha-\irm r\cos\theta}.
\end{split}
\end{equation}
With a change of the integration variable as $z=\erm^{\irm \theta}$, the last integral can be written as a contour integral along the unit circle: 
\begin{equation}
 -\frac{\partial}{\partial r} G(\rv;\alpha) 
 = \frac{1}{(2\pi)^2} \oint \frac{\drm z}{\irm z} \frac{z^2+1}{r(z^2+1) +2\irm \alpha z}
 = \frac{1}{(2\pi)^2 \irm r} \oint \drm z \frac{z^2+1}{z(z-z_+)(z-z_-)} ,
\end{equation} 
where $z_\pm=\irm (-\alpha\pm\sqrt{r^2+\alpha^2})/r$ are the locations of poles. 
Since $|z_+|<1<|z_-|$, the integral picks up the residues at $z=0$ and $z_+$, leading to 
\begin{equation}\label{eq:dGdr_Poisson}
 -\frac{\partial}{\partial r} G(\rv;\alpha) 
 =\frac1{2\pi r} \left[ \frac{1}{z_+z_-} + \frac{z_+^2+1}{z_+(z_+-z_-)} \right]
 =\frac1{2\pi r} \left( 1-\frac{\alpha}{\sqrt{r^2+\alpha^2}} \right).
\end{equation}
Therefore, $G(\zerov;\alpha)-G(\rv;\alpha)$ in Eq.\ \eqref{eq:corr_theta_G} can be calculated as
\begin{equation}\label{eq:Green_Poisson_0r}
\begin{split}
 G(\zerov;\alpha)-G(\rv;\alpha)
 &= \frac{1}{2\pi} \lim_{a_0\to 0} \int_{a_0}^r \drm r' \left( \frac{1}{r'}-\frac{\alpha}{r' \sqrt{{r'}^2+\alpha^2}} \right)\\
 &= \frac{1}{2\pi} \lim_{a_0\to 0} \left( \ln \frac{r}{a_0} +\arsinh~\frac{\alpha}{r} - \arsinh~\frac{\alpha}{a_0}\right)
 = \frac{1}{2\pi} \left( \ln \frac{r}{2\alpha} +\arsinh~\frac{\alpha}{r} \right).
\end{split}
\end{equation}
By substituting this into Eq.\ \eqref{eq:corr_theta_G} and taking the limit $r\gg \alpha$, 
we obtain Eq.\ \eqref{eq:corr_theta_r0_vorlat}. 

%%%%%%%%%%%%%%%%%%%%%%%%%%%%%%%%%%%%%%%%%%%%%
\section*{References}
%%%%%%%%%%%%%%%%%%%%%%%%%%%%%%%%%%%%%%%%%%%%%

\bibliographystyle{iopart-num}

\providecommand{\newblock}{}
\expandafter\ifx\csname url\endcsname\relax
  \def\url#1{{\tt #1}}\fi
\expandafter\ifx\csname urlprefix\endcsname\relax\def\urlprefix{URL }\fi
\providecommand{\eprint}[2][]{\url{#2}}

\bibliography{reference}
%\begin{thebibliography}{99}
%\end{thebibliography}

\end{document}